\documentclass[preprint, superscriptaddress, preprintnumbers,amsmath,amssymb]{revtex4}

\usepackage{booktabs}
\usepackage{color,graphicx}
\usepackage{amsmath}
\usepackage{array}[=2016-10-06]
\usepackage{dcolumn}
\usepackage{CJKutf8}
\usepackage{bm}                  % bold math
\usepackage{soul}
\usepackage{enumerate}
\usepackage{multirow}
\usepackage{mathrsfs}
\usepackage{braket}
\usepackage{longtable}
\usepackage{pdflscape}
\usepackage[pdfstartview=FitH,
            CJKbookmarks=true,
            bookmarksnumbered=true,
            bookmarksopen=true,
            colorlinks=true,
            pdfborder={0 0 0}, 
            linkcolor=blue,
            anchorcolor=blue,
            citecolor=blue
            ]{hyperref}

\usepackage{float}

\newcommand{\cred}{\color{red}}
\newcommand{\fra}[1]{\vspace*{0.05in} \hrule \vspace*{0.05in}
\noindent {\color{red}\textsf{F: #1}} \vspace*{0.05in}\hrule
\vspace*{0.05in}}
\newcommand{\fraadd}[1]{{\color{red} #1}}

\newcommand{\adr}[1]{\vspace*{0.05in} \hrule \vspace*{0.05in}
\noindent {\color{green}\textsf{A: #1}} \vspace*{0.05in}\hrule
\vspace*{0.05in}}
\newcommand{\adradd}[1]{{\color{green} #1}}
\newcommand{\adrst}[1]{\setstcolor{green} \st{#1}} % remove text

\newcommand{\jer}[1]{\vspace*{0.05in} \hrule \vspace*{0.05in}
\noindent {\color{blue}\textsf{J: #1}} \vspace*{0.05in}\hrule
\vspace*{0.05in}} % write a comment
\newcommand{\jerhl}[1]{{\sethlcolor{blue} \hl{#1}} } % Highlight
\newcommand{\jeradd}[1]{{\color{blue} #1}} % add text
\newcommand{\jerst}[1]{\setstcolor{blue} \st{#1}} % remove text

\begin{document}
%\begin{CJK*}{GBK}{song}
\begin{CJK*}{UTF8}{gbsn}

\title{Nuclear mass table in deformed relativistic Hartree-Bogoliubov theory in continuum, III: Nuclei with $8\leq Z \leq 120$}
% \title{Nuclear mass table in deformed relativistic Hartree-Bogoliubov theory in continuum for nuclei with $8\leq Z \leq 120$}

\author{Peng Guo}
\affiliation{State Key Laboratory of Nuclear Physics and Technology, School of Physics, Peking University, Beijing 100871, China}

\author{Xiaojie Cao}
\affiliation{School of Nuclear Science and Technology, University of Chinese Academy of Sciences, Beijing 100049, China}
\affiliation{Department of Physics, School of Science, Tianjin University, Tianjin 300354, China}

\author{Kangmin Chen}
\affiliation{School of Physics and Mechatronics Engineering, Guizhou Minzu University, Guiyang 550025, China}

\author{Qibo Chen}
\affiliation{School of Physics, East China Normal University, Shanghai 200241, China}

% \author{Zhihui Chen}
% \affiliation{Department of Physics, The University of Hong Kong, Pokfulam 999077, Hong Kong, China}

\author{Myung-Ki Cheoun}
\affiliation{Department of Physics and Origin of Matter and Evolution of Galaxy Institute, Soongsil University, Seoul 06978, Korea}

\author{Yongbeom Choi}
\affiliation{School of Physics, Peng Huanwu Collaborative Center for Research and Education, and International Research Center for Big-Bang Cosmology and Element Genesis, Beihang University, Beijing 100191, People's Republic of China}

\author{Wenmin Deng}
\affiliation{Department of Physics, Beijing Normal University, Beijing 100875, China}

\author{Jianmin Dong}
\affiliation{Institute of Modern Physics, Chinese Academy of Sciences, Lanzhou 730000, China}
\affiliation{School of Physics, University of Chinese Academy of Sciences, Beijing 100049, China}

\author{Pengxiang Du}
\affiliation{College of Physics, Jilin University, Changchun 130012, China}

\author{Xiaokai Du}
\affiliation{State Key Laboratory of Nuclear Physics and Technology, School of Physics, Peking University, Beijing 100871, China}

\author{Kangda Duan}
\affiliation{Institute of Modern Physics, Chinese Academy of Sciences, Lanzhou 730000, China}
\affiliation{School of Physics, University of Chinese Academy of Sciences, Beijing 100049, China}

\author{Xiaohua Fan}
\affiliation{School of Physical Science and Technology, Southwest University, Chongqing 400715, China}

\author{Wei Gao}
\affiliation{School of Physics, Zhengzhou University, Zhengzhou 450001, China}

\author{Lisheng Geng}
\affiliation{School of Physics, Beihang University, Beijing 102206, China}
\affiliation{School of Physics, Zhengzhou University, Zhengzhou 450001, China}

\author{Xi Guo}
\affiliation{School of Nuclear Science and Technology, Lanzhou University, Lanzhou 730000, China}

\author{Yixin Guo}
\affiliation{Center for Exotic Nuclear Studies, Institute for Basic Science, Daejeon 34126, Korea}
\affiliation{RIKEN Nishina Center, Wako 351-0198, Japan}

\author{Eunja Ha}
\affiliation{Department of Physics, Hanyang University, Seoul 04763, Korea}

\author{Xiao-Tao He}
\affiliation{College of Materials Science and Technology, Nanjing University of Aeronautics and Astronautics, Nanjing 210016, China}

\author{Jinniu Hu}
\affiliation{School of Physics, Nankai University, Tianjin 300071, China}

\author{Rongyan Hu}
\affiliation{School of Physical Science and Technology, Southwest University, Chongqing 400715, China}

\author{Jingke Huang}
\affiliation{School of Physics, Zhengzhou University, Zhengzhou 450001, China}

\author{Kun Huang}
\affiliation{College of Materials Science and Technology, Nanjing University of Aeronautics and Astronautics, Nanjing 210016, China}

\author{Yanan Huang}
\affiliation{School of Nuclear Science and Technology, Lanzhou University, Lanzhou 730000, China}

\author{Zidan Huang}
\affiliation{School of Physics, Zhengzhou University, Zhengzhou 450001, China}

% \author{Kim Da Hyung}
% \affiliation{Department of Physics, The University of Hong Kong, Pokfulam 999077, Hong Kong, China}

% \author{Chan Hoi Yat Jeffrey}
% \affiliation{Department of Physics, The University of Hong Kong, Pokfulam 999077, Hong Kong, China}

\author{Xiaofei Jiang}
\affiliation{State Key Laboratory of Nuclear Physics and Technology, School of Physics, Peking University, Beijing 100871, China}

\author{Seonghyun Kim}
\affiliation{Department of Physics and Origin of Matter and Evolution of Galaxy Institute, Soongsil University, Seoul 06978, Korea}

\author{Youngman Kim}
\affiliation{Center for Exotic Nuclear Studies, Institute for Basic Science, Daejeon 34126, Korea}

% \author{Pak Chung Lam}
% \affiliation{Department of Physics, The University of Hong Kong, Pokfulam 999077, Hong Kong, China}

\author{Chang-Hwan Lee}
\affiliation{Department of Physics, Pusan National University, Busan 46241, Korea}

\author{Jenny Lee}
\affiliation{Department of Physics, The University of Hong Kong, Pokfulam 999077, Hong Kong, China}

\author{Jian Li}
\affiliation{College of Physics, Jilin University, Changchun 130012, China}

\author{Minglong Li}
\affiliation{Department of Physics, Graduate School of Science, The University of Tokyo, Tokyo 113-0033, Japan}

\author{Zhipan Li}
\affiliation{School of Physical Science and Technology, Southwest University, Chongqing 400715, China}

\author{Zhengzheng Li}
\affiliation{State Key Laboratory of Nuclear Physics and Technology, School of Physics, Peking University, Beijing 100871, China}

\author{Zhanjiang Lian}
\affiliation{China Institute of Atomic Energy, Beijing 102413, China}

\author{Haozhao Liang}
\affiliation{Department of Physics, Graduate School of Science, The University of Tokyo, Tokyo 113-0033, Japan}
\affiliation{Quark Nuclear Science Institute, The University of Tokyo, Tokyo 113-0033, Japan}
\affiliation{RIKEN Center for Interdisciplinary Theoretical and Mathematical Sciences (iTHEMS), Wako 351-0198, Japan}

\author{Boran Liu}
\affiliation{School of Physics and Astronomy, Sun Yat-sen University, Zhuhai, 519082, China}

\author{Lang Liu}
\affiliation{School of Optoelectronic Information and Physical Science, Jiangnan University, Wuxi 214122, China}
\affiliation{Wuxi Liangxi Jiangnan Institute for Advanced Study, Wuxi 214021, China}

% \author{Zhi-Rui Liu}
% \affiliation{College of Materials Science and Technology, Nanjing University of Aeronautics and Astronautics, Nanjing 210016, China}

\author{Xiao Lu}
\affiliation{Institute of Theoretical Physics, Chinese Academy of Sciences, Beijing 100190, China}

\author{Chenjun Lv}
\affiliation{Department of Physics, Fuzhou University, Fuzhou 350108, Fujian, China}

\author{Jie Meng} \email{mengj@pku.edu.cn}
\affiliation{State Key Laboratory of Nuclear Physics and Technology, School of Physics, Peking University, Beijing 100871, China}

% \author{Ziyan Meng}
% \affiliation{State Key Laboratory of Nuclear Physics and Technology, School of Physics, Peking University, Beijing 100871, China}

\author{Myeong-Hwan Mun}
\affiliation{Department of Physics, Kyungpook National University, Daegu, 41566, Republic of Korea}

\author{Yifei Niu}
\affiliation{School of Physics and Astronomy, Shanghai Jiao Tong University, Key Laboratory for Particle Astrophysics and Cosmology (MoE), Shanghai 200240, China}
\affiliation{Shanghai Key Laboratory for Particle Physics and Cosmology, Shanghai 200240, China}

\author{Zhongming Niu}
\affiliation{School of Physics, Anhui University, Hefei 230601, China}

\author{Cong Pan}
\affiliation{Department of Physics, Anhui Normal University, Wuhu 241000, China}
\affiliation{State Key Laboratory of Nuclear Physics and Technology, School of Physics, Peking University, Beijing 100871, China}

\author{Panagiota Papakonstantinou}
\affiliation{Institute for Rare Isotope Science, Institute for Basic Science, Daejeon 34000, Korea}

\author{Jing Peng}
\affiliation{Department of Physics, Beijing Normal University, Beijing 100875, China}

\author{Xiaoying Qu}
\affiliation{School of Physics and Mechatronics Engineering, Guizhou Minzu University, Guiyang 550025, China}

\author{Tianshuai Shang}
\affiliation{College of Physics, Jilin University, Changchun 130012, China}

\author{Xinle Shang}
\affiliation{Institute of Modern Physics, Chinese Academy of Sciences, Lanzhou 730000, China}
\affiliation{School of Physics, University of Chinese Academy of Sciences, Beijing 100049, China}

\author{Caiwan Shen}
\affiliation{School of Science, Huzhou Normal University, Huzhou 313000, China}

\author{Guofang Shen}
\affiliation{Institute of Modern Physics, Chinese Academy of Sciences, Lanzhou 730000, China}

\author{Ting Sun}
\affiliation{School of Physics, Anhui University, Hefei 230601, China}

\author{Tingting Sun}
\affiliation{School of Physics, Zhengzhou University, Zhengzhou 450001, China}

\author{Wei Sun}
\affiliation{School of Physical Science and Technology, Southwest University, Chongqing 400715, China}

\author{Xiang-Xiang Sun}
\affiliation{Institute of Theoretical Physics, Chinese Academy of Sciences, Beijing 100190, China}
\affiliation{School of Nuclear Science and Technology, University of Chinese Academy of Sciences, Beijing 100049, China}

\author{Chen Wang}
\affiliation{College of Materials Science and Technology, Nanjing University of Aeronautics and Astronautics, Nanjing 210016, China}

% \author{Tianyu Wang}
% \affiliation{School of Mechatronics Engineering, Guizhou Minzu University, Guiyang 550025, China}

\author{Peng Wang}
\affiliation{School of Physics, Zhengzhou University, Zhengzhou 450001, China}

\author{Sibo Wang}
\affiliation{Department of Physics and Chongqing Key Laboratory for Strongly Coupled Physics, Chongqing University, Chongqing 401331, China}

\author{Yiran Wang}
\affiliation{Institute of High Energy Physics, Chinese Academy of Sciences, Beijing 100049, China}
\affiliation{School of Physical Sciences, University of Chinese Academy of Sciences, Beijing 100049, China}
\affiliation{Department of Physics, School of Science, Tianjin University, Tianjin 300354, China}

\author{Yuanyuan Wang}
\affiliation{Mathematics and Physics Department, North China Electric Power University, Beijing 102206, China}

\author{Yumeng Wang}
\affiliation{School of Optoelectronic Information and Physical Science, Jiangnan University, Wuxi 214122, China}
\affiliation{Wuxi Liangxi Jiangnan Institute for Advanced Study, Wuxi 214021, China}

\author{Jiawei Wu}
\affiliation{College of Materials Science and Technology, Nanjing University of Aeronautics and Astronautics, Nanjing 210016, China}

\author{Liang Wu}
\affiliation{School of Physics, Zhengzhou University, Zhengzhou 450001, China}

\author{Xinhui Wu}
\affiliation{Department of Physics, Fuzhou University, Fuzhou 350108, Fujian, China}

\author{Xuewei Xia}
\affiliation{School of Physics and Electronic Engineering, Center for Computational Sciences, Sichuan Normal University, Chengdu 610068, China}

% \author{Huihui Xie}
% \affiliation{College of Physics, Jilin University, Changchun 130012, China}

\author{Yingfeng Xu}
\affiliation{Department of Physics, The University of Hong Kong, Pokfulam 999077, Hong Kong, China}

\author{Yonghe Xue}
\affiliation{Institute of Modern Physics, Chinese Academy of Sciences, Lanzhou 730000, China}
\affiliation{School of Physics, University of Chinese Academy of Sciences, Beijing 100049, China}

\author{Jiangming Yao}
\affiliation{School of Physics and Astronomy, Sun Yat-sen University, Zhuhai, 519082, China}
\affiliation{Facility for Rare Isotope Beams, Michigan State University, East Lansing MI, 48824-1321, USA}

\author{Yaochen Yang}
\affiliation{State Key Laboratory of Nuclear Physics and Technology, School of Physics, Peking University, Beijing 100871, China}

% \author{Tammi Ip Kwan Yau}
% \affiliation{Department of Physics, The University of Hong Kong, Pokfulam 999077, Hong Kong, China}

\author{Guixiang Ye}
\affiliation{National Key Laboratory of Neutron Science and Technology, Institute of Nuclear Physics and Chemistry, China Academy of Engineering Physics, Mianyang, Sichuan 621900, China}

\author{Lian Ye}
\affiliation{School of Physics and Mechatronics Engineering, Guizhou Minzu University, Guiyang 550025, China}

\author{To Chung Yiu}
\affiliation{School of Physical Science and Technology, Southwest University, Chongqing 400715, China}

\author{Jianghan Yu}
\affiliation{School of Science, Huzhou Normal University, Huzhou 313000, China}

\author{Yangyang Yu}
\affiliation{School of Physics and Mechatronics Engineering, Guizhou Minzu University, Guiyang 550025, China}

\author{Wentao Zeng}
\affiliation{State Key Laboratory of Nuclear Physics and Technology, School of Physics, Peking University, Beijing 100871, China}
\affiliation{Department of Physics, School of Science, Tianjin University, Tianjin 300354, China}

\author{Kaiyuan Zhang}
\affiliation{National Key Laboratory of Neutron Science and Technology, Institute of Nuclear Physics and Chemistry, China Academy of Engineering Physics, Mianyang, Sichuan 621900, China}
\affiliation{State Key Laboratory of Nuclear Physics and Technology, School of Physics, Peking University, Beijing 100871, China}

\author{Jinzhe Zhang}
\affiliation{School of Physics, Nankai University, Tianjin 300071, China}

\author{Ning Zhang}
\affiliation{College of Materials Science and Technology, Nanjing University of Aeronautics and Astronautics, Nanjing 210016, China}

\author{Shijie Zhang}
\affiliation{School of Physics, Zhengzhou University, Zhengzhou 450001, China}

\author{Shuangquan Zhang}
\affiliation{State Key Laboratory of Nuclear Physics and Technology, School of Physics, Peking University, Beijing 100871, China}

\author{Wei Zhang}
\affiliation{School of Physics, Zhengzhou University, Zhengzhou 450001, China}

\author{Xiaoyan Zhang}
\affiliation{School of Physics, Anhui University, Hefei 230601, China}

\author{Yanxin Zhang}
\affiliation{School of Physics and Astronomy, Sun Yat-sen University, Zhuhai, 519082, China}

\author{Ying Zhang}
\affiliation{Department of Physics, School of Science, Tianjin University, Tianjin 300354, China}

\author{Yingxun Zhang}
\affiliation{China Institute of Atomic Energy, Beijing 102413, China}
\affiliation{Guangxi Key Laboratory of Nuclear Physics and Technology, Guangxi Normal University, Guilin 541004, China}

\author{Zhenhua Zhang}
\affiliation{Mathematics and Physics Department, North China Electric Power University, Beijing 102206, China}

\author{Qiang Zhao}
\affiliation{China Institute of Atomic Energy, Beijing 102413, China}

% \author{Yingchun Zhao}
% \affiliation{State Key Laboratory of Nuclear Physics and Technology, School of Physics, Peking University, Beijing 100871, China}

\author{Ruyou Zheng}
\affiliation{School of Physics, Beihang University, Beijing 102206, China}

\author{Chang Zhou}
\affiliation{State Key Laboratory of Nuclear Physics and Technology, School of Physics, Peking University, Beijing 100871, China}

\author{Shan-Gui Zhou}
\affiliation{Institute of Theoretical Physics, Chinese Academy of Sciences, Beijing 100190, China}
\affiliation{School of Physical Sciences, University of Chinese Academy of Sciences, Beijing 100049, China}
\affiliation{School of Nuclear Science and Technology, University of Chinese Academy of Sciences, Beijing 100049, China}
\affiliation{School of Physics, Peng Huanwu Collaborative Center for Research and Education, and International Research Center for Big-Bang Cosmology and Element Genesis, Beihang University, Beijing 100191, People's Republic of China}

\author{Lianjian Zou}
\affiliation{School of Physics, Nankai University, Tianjin 300071, China}

\collaboration{DRHBc Mass Table Collaboration}

\begin{abstract}
The mass table in the deformed relativistic Hartree–Bogoliubov theory in continuum (DRHBc) with the PC-PK1 density functional has been established for nuclei with $8 \leq Z \leq 120$, extended from the previous works for even–even nuclei~\href{https://doi.org/10.1016/j.adt.2022.101488}{(Zhang et al. (DRHBc mass table collaboration), At. Data Nucl. Data Tables 144, 101488 (2022))} and for even-$Z$ nuclei~\href{https://doi.org/10.1016/j.adt.2024.101661}{(Guo et al. (DRHBc mass table collaboration), At. Data Nucl. Data Tables 158, 101661 (2024))}.
The calculated binding energies, two- and one-nucleon separation energies, root-mean-square (rms) radii of neutron, proton, matter, and charge distributions, quadrupole deformations, neutron and proton Fermi surfaces, and the blocked neutron (proton) orbitals of odd-$N$ ($Z$) nuclei are tabulated and compared with the available experimental data. 
A total of 9495 nuclei are predicted to be bound, with an rms deviation of 1.444 MeV from the 2380 mass data.
Good agreement with the available experimental pairing gaps, $\alpha$ decay energies, and charge radii is also achieved. The accuracies of the calculated nuclear masses and nucleon separation energies as well as the prediction for drip lines are compared with those obtained by other relativistic and nonrelativistic density functional calculations. It turns out that the DRHBc theory with PC-PK1 provides one of the best microscopic descriptions for nuclear masses. The systematics of nucleon separation energies, pairing gaps, pairing energies, two-nucleon gaps, $\alpha$ decay energies, rms radii, quadrupole deformations, potential energy curves, neutron density distributions, and neutron mean-field potentials are discussed. 
\end{abstract}

\date{\today}

\maketitle
\tableofcontents

%%%%%%%%%%%%%%%%%%%%%%%%%%%%%%%%%%%%%%%%%%%%%%%%%%%%%%%%%%
%                    begin  introduction
%%%%%%%%%%%%%%%%%%%%%%%%%%%%%%%%%%%%%%%%%%%%%%%%%%%%%%%%%%

%----------------------------------------------------------------------------------------
\section{Introduction}
%----------------------------------------------------------------------------------------

In the previous works~\cite{Zhang2022mass,Guo2024mass}, following the strategies and techniques presented in Refs.~\cite{Zhang2020PRC,Pan2022PRC}, the nuclear mass table calculated by the deformed relativistic Hartree-Bogoliubov theory in continuum (DRHBc)~\cite{Zhou2010PRC,Li2012PRC,Li2012CPL} with the density functional PC-PK1~\cite{Zhao2010PRC} was constructed for even-even nuclei~\cite{Zhang2022mass} and even-$Z$ nuclei~\cite{Guo2024mass}.
A total of 2583 even-even nuclei~\cite{Zhang2022mass} and 2246 even-$Z$ odd-$N$~\cite{Guo2024mass} with $8\le Z\le 120$ have been predicted to be bound.
The root-mean-square (rms) deviation for 1244 nuclei with mass measured is 1.433 MeV, providing one of the best density functional descriptions for nuclear masses~\cite{Guo2024mass}.
With the completion of DRHBc calculations for odd-$Z$ nuclei, this paper reports the DRHBc mass table for all the nuclei with $8 \le Z \le 120$.

The DRHBc theory, based on meson-exchange~\cite{Zhou2010PRC,Li2012PRC,Li2012CPL,Chen2012PRC} or point-coupling~\cite{Zhang2020PRC,Pan2022PRC} density functionals, solves the deformed relativistic Hartree-Bogoliubov (RHB) equations in a Dirac Woods-Saxon (DWS) basis~\cite{Zhou2003PRC}. In this way, the deformation, pairing, and continuum effects are self-consistently incorporated, and thus a microscopic description for both exotic and stable nuclei is provided.
% The DRHBc theory, based on meson-exchange~\cite{Zhou2010PRC,Li2012PRC,Li2012CPL,Chen2012PRC} or point-coupling~\cite{Zhang2020PRC,Pan2022PRC} density functionals, where the deformed relativistic Hartree-Bogoliubov (RHB) equations are solved in a Dirac Woods-Saxon (DWS) basis~\cite{Zhou2003PRC}, including the deformation degrees of freedom, pairing correlations, and continuum effects in a self-consistent way, provides a microscopic description for both exotic and stable nuclei.
The DRHBc theory has described successfully the halo structures in $^{17}$B~\cite{Yang2021PRL}, $^{19}$B~\cite{Sun2021PRC}, $^{15,19,22}$C~\cite{SUN2018PLB,SUN2020NPA,Wang2024EPJA}, $^{12}$N~\cite{ZhangKY2025PLB}, $^{29}$F~\cite{Zhang2026Par}, $^{31}$Ne~\cite{Zhong2022SCP,Pan2024PLB}, and $^{37}$Mg~\cite{Zhang2023PLB,An2024PLB,Zhang2024NPR}, predicted the possible halo in $^{31}$F~\cite{Zhang2026Par}, $^{39,41}$Na~\cite{Zhang2023PRC}, $^{42,44}$Mg~\cite{Zhou2010PRC,Li2012PRC}, and $^{43,45}$Si~\cite{Pan2026PLB}, and explored the possibility of a one-proton halo in $^{22}$Al~\cite{Zhang2024PRC,Pana2025PRC}.
In addition to the halo phenomena, it has also succeeded in describing many other nuclear phenomena and properties, such as the prolate-shape dominance in Te, Xe, and Ba isotopes~\cite{Guo2023PRC}, the shape evolution of Ho isotopes~\cite{ZhangJZ2026PRC}, the $\alpha$ decay half-lives for even-even nuclei from W to U~\cite{Choi2024PRC}, the shape evolution of neutron-deficient Au and kink structure of charge radii in Pb isotopes~\cite{MunHa2024PRC}, the odd-even shape staggering and the charge-radius kink structure in Hg isotopes~\cite{Mun2023PLB}, the neutron skin thickness in Pb isotopes~\cite{Kim2022PRC}, the inner fission barrier of U isotopes~\cite{ZhangW2024CPC}, the charge radii and their correlations with quadrupole deformation~\cite{CongP2025PRC}, and newly measured mass data~\cite{Qu2025NST}.

To probe novel structures or mechanisms in nuclei, the DRHBc theory has been used to investigate the nucleon in the classically forbidden region in Mg isotopes~\cite{ZhangK2019PRC}, the deformation effects on the drip-line locations~\cite{EJ2021IJMPE}, the possible peninsula of stability~\cite{Zhang2021PRC,Pan2021PRC,He2021CPC,He2024PRC}, the possible shape coexistence and bubble structure~\cite{In2020JKPS,Choi2022PRC,Mun2025Par,Song2025Par}, the prolate-oblate shape competition and the volume/surface contribution to neutron skin thickness of Bk isotopes~\cite{Huang2025PRC,Wang2026PRC,Sun2026NST}, the evolution of $N = 20,28$, and 50 shell closures in the $20 \leq Z \leq 30$ region~\cite{Zheng2024CPC}, the double magicity of $^{164}$Pb~\cite{ZhangPan2025PRC}, possible magic numbers in the superheavy mass region~\cite{ZhangXY2024PRC,Liu2024Par,Du2024Par,Pan2025Par,Zhuang2026CPC} and the influence of bubble structure~\cite{ZhangXY2024PRC} and high-angular-momentum single-particle states~\cite{Wang2024Par}, the symmetry energy in relation with two-nucleon separation energies of Ca and Pb isotopes~\cite{Mun2024PRC}, the interdependence of pairing energy and mean field energy in reaching the energy minimum~\cite{Mun2025PRC}, and the sensitivity of $r$-process abundances to nuclear deformation~\cite{Pan2025EPJA,Choi2026PRC}.

Based on the DRHBc theory, significant extensions and optimizations have been done and reviewed in Ref.~\cite{Zhang2025AAPPS}.
The angular momentum projection has been implemented to study the rotational excitations of exotic nuclei~\cite{SunZhou2021PRC,Sun2021SB,Sun2024NPR}.
The finite amplitude method has been implemented to investigate the monopole and dipole resonances in exotic nuclei ~\cite{Sun2022PRC,Jiang2026PLB}.
Two-dimensional collective Hamiltonian (2DCH) method has been implemented to study the beyond mean-field dynamical correlations and improve the descriptions of ground-state properties~\cite{Sun2022CPC,ZhangXY2023PRC}.
The interaction cross section, longitudinal momentum distribution, and the charged-particle evaporation process contribution to the charge-changing cross section have been investigated based on the Glauber model combined with the structure inputs from DRHBc calculations~\cite{Zhong2022SCP,Zhao2023PLB,Wang2024EPJA,Zhao2024PLB,An2024PLB}.
The one-proton emission in $^{149-151}$Lu has been studied based on the WKB model combined with the structure properties from DRHBc calculations~\cite{Xiao2023PLB,Lu2024PLB,Fan2026PRC}.
The DRHBc theory has been improved by optimizing the Dirac Woods-Saxon basis~\cite{ZaP2022PRC}, implementing the Lipkin-Nogami method~\cite{Zhou2024Par,Zhou2026PRC}, and including the time-odd component~\cite{Pan2024PLB}.
The DRHBc theory has been extended to include the triaxial degree of freedom, $\textit{i.e.}$, the triaxial relativistic Hartree-Bogoliubov theory in continuum (TRHBc)~\cite{Zhang2023PRCL,Xiang2023Sym,ZhangTRHBc2025PRC}.
In Refs.~\cite{Wu2024PRC,Wu2024PRC_2}, the correction of the mass table in relativistic continuum Hartree-Bogoliubov (RCHB) theory~\cite{Xia2018ADNDT} has been investigated by machine learning and compared with the DRHBc mass table.

% The above achievements demonstrate the DRHBc theory as a powerful tool for studying both stable and exotic nuclei, and the comprehensive DRHBc mass table has the potential to provide a high-precision description for nuclear masses.
In this paper, we present the DRHBc mass table for nuclei with $8\le Z\le120$.
The theoretical framework of DRHBc is briefly introduced in Sec.~\ref{theory}.
Numerical details are listed in Sec.~\ref{numerical}. 
The results and discussions are compiled in Sec.~\ref{results}, including nuclear masses, nucleon separation energies, pairing gaps, pairing energies, two-nucleon gaps, alpha-decay energies, rms radii, quadrupole deformations, potential energy curves, neutron density distributions, and neutron potentials.
Finally, a summary is given in Sec.~\ref{summary}.

%%%%%%%%%%%%%%%%%%%%%%%%%%%%%%%%%%%%%%%%%%%%%%%%%%%%%%%%%%
%                    begin  theoretical framework
%%%%%%%%%%%%%%%%%%%%%%%%%%%%%%%%%%%%%%%%%%%%%%%%%%%%%%%%%%

%----------------------------------------------------------------------------------------
\section{Theoretical framework} \label{theory}
%----------------------------------------------------------------------------------------

The details of the DRHBc theory with meson-exchange and point-coupling density functionals can be found in Ref.~\cite{Li2012PRC} and Refs.~\cite{Zhang2020PRC,Pan2022PRC}, respectively. In the following, the theoretical framework of the DRHBc theory will be briefly introduced.

The relativistic density functional theory starts from an effective Lagrangian with either the meson-exchange or point-coupling interaction~\cite{Meng2016RDFNS}. For the nonlinear point-coupling interaction adopted here, the Lagrangian reads
\begin{align}\label{Lagrangian}
\mathscr{L} = & \bar{\psi}(i\gamma_\mu\partial^\mu-M)\psi - \frac{1}{2}\alpha_S(\bar{\psi}\psi)(\bar{\psi}\psi)\notag \\
& -\frac{1}{2}\alpha_V(\bar{\psi}\gamma_\mu\psi)(\bar{\psi}\gamma^\mu\psi)-\frac{1}{2}\alpha_{TV}(\bar{\psi}\vec{\tau}\gamma_\mu\psi)(\bar{\psi}\vec{\tau}\gamma^\mu\psi)\notag \\
& -\frac{1}{3}\beta_S(\bar{\psi}\psi)^3-\frac{1}{4}\gamma_S(\bar{\psi}\psi)^4-\frac{1}{4}\gamma_V[(\bar{\psi}\gamma_\mu\psi)(\bar{\psi}\gamma^\mu\psi)]^2\notag \\
&-\frac{1}{2}\delta_S\partial_\nu(\bar{\psi}\psi)\partial^\nu(\bar{\psi}\psi)-\frac{1}{2}\delta_V\partial_\nu(\bar{\psi}\gamma_\mu\psi)\partial^\nu(\bar{\psi}\gamma^\mu\psi)\notag \\
&-\frac{1}{2}\delta_{TV}\partial_\nu(\bar{\psi}\vec{\tau}\gamma_\mu\psi)\partial^\nu(\bar{\psi}\vec{\tau}\gamma^\mu\psi)\notag \\
&-\frac{1}{4}F^{\mu\nu}F_{\mu\nu}-e\bar{\psi}\gamma^{\mu}\frac{1-\tau_3}{2} {\psi} A_{\mu},
\end{align}
where $M$ is the nucleon mass, $e$ is the charge unit, and $A_{\mu}$ and $F_{\mu \nu}$ are the four-vector potential and field strength tensor of the electromagnetic field, respectively. Here $\alpha_S$, $\alpha_V$, and $\alpha_{TV}$ represent the coupling constants for four-fermion terms, $\beta_S$, $\gamma_S$, and $\gamma_V$ are those for the high-order terms, and $\delta_S$, $\delta_V$, and $\delta_{TV}$ refer to those for gradient terms. The subscripts $S$, $V$, and $TV$ denote scalar, vector, and isovector-vector channels, respectively. The isovector-scalar channel is neglected since its inclusion does not necessarily improve the description of nuclear ground-state properties~\cite{TB2002PRC}.

The energy density functional can then be constructed from the Lagrangian density of Eq.~(\ref{Lagrangian}) under the mean-field and no-sea approximations. 
By minimizing the energy density functional with respect to the densities, the Dirac equation for nucleons can be obtained~\cite{Meng2016RDFNS}.
Treating the mean field and pairing correlations self-consistently, the relativistic Hartree-Bogoliubov equation reads~\cite{Kucharek1991ZPA},  
\begin{equation}\label{RHB}
\left(\begin{matrix}
\hat{h}_D-\lambda_\tau & \hat{\Delta} \\
-\hat{\Delta}^* &-\hat{h}_D^*+\lambda_\tau
\end{matrix}\right)\left(\begin{matrix}
U_k\\
V_k
\end{matrix}\right)=E_k\left(\begin{matrix}
U_k\\
V_k
\end{matrix}\right),
\end{equation}
where $(U_k,V_k)^\text{T}$ are the quasiparticle wave functions, $E_k$ is the quasiparticle energy, and $\lambda_\tau$ is the Fermi surface ($\tau = \mathrm{n}/\mathrm{p}$ for neutrons or protons).
The Dirac Hamiltonian $h_D$ in coordinate space reads 
\begin{equation}
h_D(\bm{r})=\bm{\alpha}\cdot\bm{p}+V(\bm{r})+\beta[M+S(\bm{r})],
\end{equation}
with the scalar $S(\bm r)$ and vector $V(\bm r)$ potentials
\begin{align}
S(\bm r) =& \alpha_S \rho_S + \beta_S \rho^2_S + \gamma_S \rho^3_S + \delta_S \Delta \rho_S, \label{P_S}\\
V(\bm r) =& \alpha_V \rho_V + \gamma_V \rho^3_V + \delta_V \Delta \rho_V + eA^0 + \alpha_{TV}\tau_3\rho_3 + \delta_{TV} \tau_3 \Delta \rho_3,\label{P_V}
\end{align}
constructed by the local densities:
\begin{align}
\rho_S(\bm r) =& \sum_{k>0} V^\dag_k (\bm r) \gamma_0 V_k (\bm r),\\
\rho_V(\bm r) =& \sum_{k>0} V^\dag_k (\bm r) V_k (\bm r),\\
\rho_3(\bm r) =& \sum_{k>0} V^\dag_k (\bm r) \tau_3 V_k (\bm r).
\end{align}
% \begin{equation}
% \rho_S(\bm r) = \sum_{k>0} V^\dag_k (\bm r) \gamma_0 V_k (\bm r),
% \end{equation}
% \begin{equation}
% \rho_V(\bm r) = \sum_{k>0} V^\dag_k (\bm r) V_k (\bm r),
% \end{equation}
% \begin{equation}
% \rho_3(\bm r) = \sum_{k>0} V^\dag_k (\bm r) \tau_3 V_k (\bm r).
% \end{equation}
The summation runs over the quasiparticle states with positive energies.

The pairing potential $\Delta$ reads 
\begin{equation}\label{Delta}
\Delta(\bm r_1,\bm r_2) = V^{\mathrm{pp}}(\bm r_1,\bm r_2)\kappa(\bm r_1,\bm r_2),
\end{equation}
with the density-dependent zero-range pairing force,
\begin{equation}\label{pair}
V^{\mathrm{pp}}(\bm r_1,\bm r_2)= V_0 \frac{1}{2}(1-P^\sigma)\delta(\bm r_1-\bm r_2)\left(1-\frac{\rho(\bm r_1)}{\rho_{\mathrm{sat}}}\right),
\end{equation}
and the pairing tensor,
\begin{equation}
\kappa(\bm r_1,\bm r_2) = \sum_{k>0} V_k^* (\bm r_1) U_k^\text{T} (\bm r_2).
\end{equation}

For axially deformed nuclei, the potentials and densities are expanded by the Legendre polynomials, 
\begin{equation}\label{legendre}
f(\bm r)=\sum_{\lambda} f_{\lambda}(r)P_{\lambda}(\cos\theta),~~\lambda=0,2,4,\cdots,
\end{equation}
% where $\lambda$ is the order of a Legendre polynomial $P_\lambda$, and $f_\lambda$ is the corresponding radial component calculated from
with  
\begin{equation}\label{leg_f}
f_{\lambda}(r)=\frac{2\lambda+1}{4\pi}\int \mathrm{d}\Omega f(\bm{r}) P_\lambda (\cos \theta).
\end{equation}
The assumption of spatial reflection symmetry restricts the Legendre polynomials to even orders.
%If the triaxial deformation degree of freedom is included, the potentials and densities should be expanded by spherical harmonics~\cite{Zhang2023PRCL}.

For exotic nuclei with Fermi surfaces close to the continuum threshold, the continuum effects play an important role. To include the continuum effects, the deformed RHB equations are solved in a spherical DWS basis in coordinate space~\cite{Zhou2003PRC}, 
\begin{equation}\label{DWS}
    \phi_{n\kappa m}(\bm{r}s) = \frac{1}{r}\begin{pmatrix} i G_{n\kappa}(r)Y^l_{jm}(\Omega s) \\ -F_{n\kappa}(r)Y^{\tilde{l}}_{jm}(\Omega s)
    \end{pmatrix},
\end{equation}
derived from the Dirac equation with a Woods-Saxon potential. In Eq.~\eqref{DWS}, $G_{n\kappa}(r)/r$ and $F_{n\kappa}(r)/r$ are the radial wave functions for the large and small components, and $Y^{l(\tilde{l})}_{jm}(\Omega s)$ denotes spinor spherical harmonics, where $n$ is the radial quantum number, $\kappa = (-1)^{j+l+1/2}(j+1/2)$, and $\tilde{l}=l+(-1)^{j+l-1/2}$. The radial wave functions of the DWS basis have a proper asymptotic behavior at large $r$, enabling the DRHBc theory to properly account for the contributions of the continuum and describe the exotic nuclei with large spatial extensions. Due to the spatial reflection symmetry and axial symmetry, the parity $\pi$ and the third component of the angular momentum $m$ are good quantum numbers. The RHB matrix can then be decomposed into separate $m^{\pi}$ blocks, greatly simplifying the diagonalization. The resulting quasiparticle wave functions are used to construct new densities and then new potentials, which are iterated in the RHB equations until convergence is achieved.

For an odd-$A$ or odd-odd nucleus, the blocking effect of the unpaired nucleon(s) must be taken into account~\cite{NMBP1980}. The blocked orbital(s) will break the time reversal symmetry, leading to the emergence of nucleon currents.
This will increase the computational cost, especially for large-scale calculations. Therefore, the equal filling approximation (EFA)~\cite{Martin2008PRC} is adopted to save computational resources~\cite{Li2012CPL,Pan2022PRC}, which is expected to reproduce the energies of the ground state and low-lying excited states around 50 keV on average in comparison with the exact blocking procedure~\cite{Schunck2010PRC}.

Due to the mean-field approximation, the translational and rotational invariances are violated in the DRHBc theory. The broken symmetry can be restored by beyond-mean-field approaches such as the angular momentum projection~\cite{Niksic2011PPNP}. In the present DRHBc calculations, beyond-mean-field correlations are included through the microscopic calculation of center-of-mass and rotational correction energies, and the details can be found in Refs.~\cite{Zhang2020PRC,Guo2024mass}.

%%%%%%%%%%%%%%%%%%%%%%%%%%%%%%%%%%%%%%%%%%%%%%%%%%%%%%%%%%%
%%                    begin  numerical details
%%%%%%%%%%%%%%%%%%%%%%%%%%%%%%%%%%%%%%%%%%%%%%%%%%%%%%%%%%%

%%----------------------------------------------------------------------------------------
\section{Numerical details}\label{numerical}
%%----------------------------------------------------------------------------------------

The numerical details for the DRHBc mass table calculations have been investigated in Refs.~\cite{Zhang2020PRC,Pan2022PRC}. For completeness, in the following we summarize the numerical details. 

\begin{itemize}
  \item [$\bullet$] The density functional PC-PK1~\cite{Zhao2010PRC}, which provides one of the best density-functional descriptions for nuclear masses~\cite{Zhao2012PRCmass,Zhang2014FP,Lu2015PRC,Zhang2021PRC,Qu2025NST}, is employed.
  \item [$\bullet$] In the pairing channel, the density-dependent zero-range pairing force in Eq.~(\ref{pair}) is adopted, where the pairing strength $V_0=-325~\mathrm{MeV~fm}^3$ and the saturation density $\rho_{\mathrm{sat}}=0.152~\mathrm{fm}^{-3}$ together with a pairing window of $100$ MeV. With these, the odd-even mass differences for calcium and lead isotopes are well reproduced~\cite{Zhang2020PRC}.
  \item [$\bullet$] The energy cutoff for the DWS basis in the Fermi sea is set to $E^+_{\mathrm{cut}}=300$ MeV and the number of states in the Dirac sea is the same as that in the Fermi sea, which guarantees the convergence accuracy of 0.01~MeV for the total energies of doubly-magic nuclei $^{40}$Ca, $^{100}$Sn, and $^{208}$Pb~\cite{Zhang2020PRC}, and the deformed heavy odd-$A$ nucleus $^{301}$Th~\cite{Pan2022PRC}.
  % \item [$\bullet$] The number of DWS basis states in the Dirac sea is the  as that in the Fermi sea~\cite{Zhou2003PRC,Zhou2010PRC,Li2012PRC}.
  \item [$\bullet$] The angular momentum cutoff for the DWS basis $J_{\max}=23/2~\hbar$, which guarantees the convergence accuracy of $0.01\%$ for the total energies of the deformed heavy nuclei $^{300,301}$Th~\cite{Zhang2020PRC,Pan2022PRC}.
  % \item [$\bullet$] The Legendre expansion truncations in Eq.~(\ref{legendre}) are chosen as $\lambda_{\max}=6$ and $8$ for nuclei with $8\le Z\le 70$ and $71\le Z\le 100$, respectively~\cite{Zhang2020PRC,Zhang2021PRC,Pan2019IJMPE}. This guarantees the convergence accuracy of $0.03\%$ for the total energies of $^{20}$Ne and $^{112}$Mo, and of $0.01\%$ for the total energies of $^{300,301}$Th, when their deformations are constrained to be $\beta_2 = 0.6$~\cite{Zhang2020PRC,Pan2022PRC}. For superheavy nuclei with $101\le Z\le 120$, $\lambda_{\max}=10$ is adopted~\cite{Zhang2022mass,Guo2024mass}.
  \item [$\bullet$] In Refs.~\cite{Zhang2020PRC,Pan2022PRC}, the Legendre expansion truncations in Eq.~(\ref{legendre}) are chosen as $\lambda_{\max}=6$ and $8$ for nuclei with $8\le Z\le 70$ and $Z\geq 71$, respectively. This guarantees the convergence accuracy for the total energies of $0.03\%$ of $^{20}$Ne and $^{112}$Mo, and of $0.01\%$ of $^{300,301}$Th, when their deformations are constrained as $\beta_2 = 0.6$. In the calculations for the mass table~\cite{Zhang2022mass,Guo2024mass,Pan2022PRC}, $\lambda_{\max}=10$ is adopted to guarantee the same accuracy for superheavy nuclei with $101\le Z\le 120$.
  \item [$\bullet$] For odd-$A$ and odd-odd nuclei, the automatic blocking procedure is implemented~\cite{Pan2022PRC}, in which the orbital with the lowest quasiparticle energy is blocked during the iteration. If the automatic blocking procedure doesn't work, the orbitals near the Fermi surface will be blocked independently, and the result with the lowest total energy is identified as the ground state. 
\end{itemize}
%%%%%%%%%%%%%%%%%%%%%%%%%%%%%%%%%%%%%%%%%%%%%%%%%%%%%%%%%%
%                    begin  results and discussions
%%%%%%%%%%%%%%%%%%%%%%%%%%%%%%%%%%%%%%%%%%%%%%%%%%%%%%%%%%

%----------------------------------------------------------------------------------------
\section{Results and discussion}\label{results}

Systematic calculations have been performed for all odd-$Z$ nuclei from $Z = 8$ to $Z = 120$ from the proton drip line to the neutron drip line. Together with the results for even-$Z$ nuclei compiled in Refs.~\cite{Zhang2022mass,Guo2024mass}, the ground-state properties of all nuclei from $Z = 8$ to $Z = 120$ in DRHBc are summarized in Table~\ref{tab1}.
The mass number $A$, neutron number $N$, binding energy $E_{\mathrm{b}}^{\mathrm{cal}}$, binding energy including rotational correction energy $E^{\mathrm{cal}}_{\mathrm{b}+\mathrm{rot}}$, two-neutron separation energy $S_\mathrm{2n}$, two-proton separation energy $S_\mathrm{2p}$, one-neutron separation energy $S_\mathrm{n}$, one-proton separation energy $S_\mathrm{p}$, neutron rms radius $R_\mathrm{n}$, proton rms radius $R_\mathrm{p}$, matter rms radius $R_\mathrm{m}$, charge radius $R_{\mathrm{ch}}$, neutron quadrupole deformation $\beta_\mathrm{2n}$, proton quadrupole deformation $\beta_\mathrm{2p}$, total quadrupole deformation $\beta_{2}$, neutron Fermi surface $\lambda_\mathrm{n}$, and proton Fermi surface $\lambda_\mathrm{p}$ are listed.
For the odd-$A$ and odd-odd nuclei, the blocked neutron orbital $m^\pi (N)$ and proton orbital $m^\pi (P)$ or both are tabulated. 
The available data for binding energies~\cite{AME2020(1),AME2020(2),AME2020(3)} and charge radii~\cite{Angeli2013ADNDT,Li2021ADNDT} are also listed for comparison. 
In each isotopic chain, all nuclei from the proton drip line to the neutron drip line are listed in Table~\ref{tab1}. In addition, two neutrons more and less unbound nuclei beyond the drip lines are also included. 
The physical quantities for unbound nuclei with the negative separation energies or the positive Fermi surfaces are underlined for clarity.

%----------------------------------------------------------------------------------------
\subsection{Nuclear masses}
\begin{figure}[htbp]
  \centering
  \includegraphics[width=1.0\textwidth]{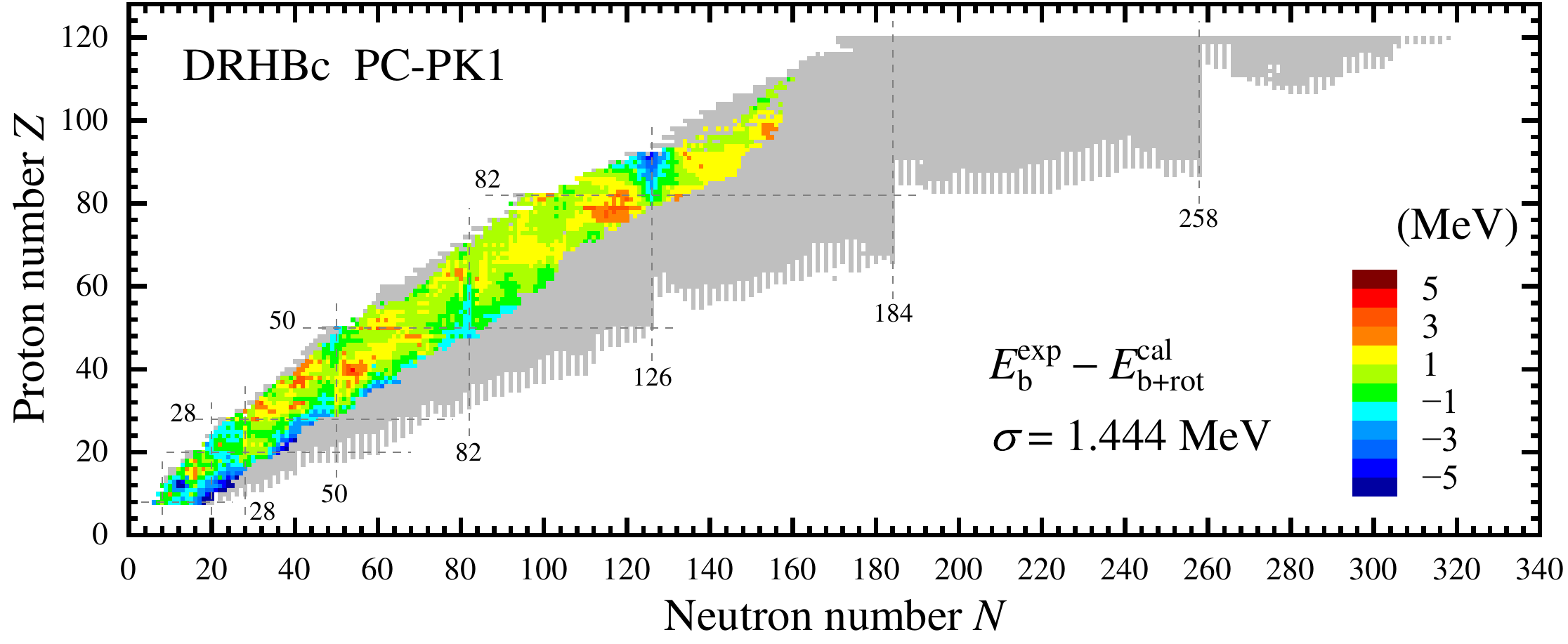}
  \caption{(Color online) 9495 bound nuclei from O ($Z = 8$) to $Z = 120$ predicted by the DRHBc theory with PC-PK1. For the 2380 nuclei with the mass measured, the binding energy differences between the data~\cite{AME2020(3)} and the DRHBc calculations (with rotational correction energy included) are scaled by colors, with the corresponding rms deviation $\sigma = 1.444$~MeV.}
\label{fig1}
\end{figure}
%----------------------------------------------------------------------------------------

Figure~\ref{fig1} illustrates the obtained nuclear landscape from $Z = 8$ to $Z = 120$. The binding energy differences between the experimental values $E^{\mathrm{exp}}_{\mathrm{b}}$ and the calculated ones $E^{\mathrm{cal}}_{\mathrm{b}+\mathrm{rot}}$ are scaled by colors. 
There are in total 9495 nuclei predicted to be bound.
The DRHBc mass table provides a satisfactory overall description for the nuclear binding energies.
% Some nuclei lying beyond the primary neutron drip lines are predicted to be bound, forming peninsulas extending from the mainland, which will be discussed in detail later in Sec.~\ref{sec:lim}.
% Among 9495 bound nuclei, the mass data of 2380 nuclei ( even-even,  even-odd,  odd-even, and  odd-odd nuclei) are available.
For the 2380 bound nuclei with mass measured, the rms deviation $\sigma=1.444$ MeV. 
For 634 even-even, 610 even-odd, 573 odd-even, and 563 odd-odd nuclei, the rms deviations are 1.518 MeV, 1.363 MeV, 1.495 MeV, and 1.424 MeV, respectively.

%----------------------------------------------------------------------------------------
\subsection{Nucleon separation energies}\label{tns}
%----------------------------------------------------------------------------------------

% Nucleon separation energies are important quantities which provide information on the nuclear stability against nucleon emission and reflect nuclear shell structure~\cite{Bender2003RMD,Sorlin2008PPNP}.
% The two-neutron, two-proton, one-neutron, one-proton, multi-neutron, and multi-proton separation energies are defined as follows,
% \begin{gather}
% S_\mathrm{2n}(Z,N) = E_{\mathrm{b}}(Z,N) - E_{\mathrm{b}}(Z,N-2),\\
% S_\mathrm{2p}(Z,N) = E_{\mathrm{b}}(Z,N) - E_{\mathrm{b}}(Z-2,N),\\
% S_\mathrm{n}(Z,N) = E_{\mathrm{b}}(Z,N) - E_{\mathrm{b}}(Z,N-1),\\
% S_\mathrm{p}(Z,N) = E_{\mathrm{b}}(Z,N) - E_{\mathrm{b}}(Z-1,N),\\
% S_{X\mathrm{n}}(Z,N)= E_{\mathrm{b}}(Z,N) - E_{\mathrm{b}}(Z,N-X), X\geq 3\\
% S_{Y\mathrm{p}}(Z,N) = E_{\mathrm{b}}(Z,N) - E_{\mathrm{b}}(Z-Y,N), Y\geq 3
% \end{gather}

Nucleon separation energies are important quantities to provide information on the nuclear stability against nucleon emission and reflect nuclear shell structure~\cite{Bender2003RMD,Sorlin2008PPNP}. The definitions are
\begin{gather}
S_{x\mathrm{n}}(Z,N)= E_{\mathrm{b}}(Z,N) - E_{\mathrm{b}}(Z,N-x), \\
S_{x\mathrm{p}}(Z,N) = E_{\mathrm{b}}(Z,N) - E_{\mathrm{b}}(Z-x,N). 
\end{gather}
Here, $x = 1$ corresponds to the one-nucleon separation energies, $x = 2$ to the two-nucleon separation energies, and $ x \geq 3$ to the multi-nucleon separation energies.
In this paper, a nucleus is considered bound only if its one-nucleon, two-nucleon, and multi-nucleon separation energies are all positive and its neutron and proton Fermi surfaces are negative.
For each isotopic chain, the heaviest and lightest bound nuclei define the neutron and proton drip lines, respectively.

\begin{figure}[htbp]
  \centering
  \includegraphics[width=1.0\textwidth]{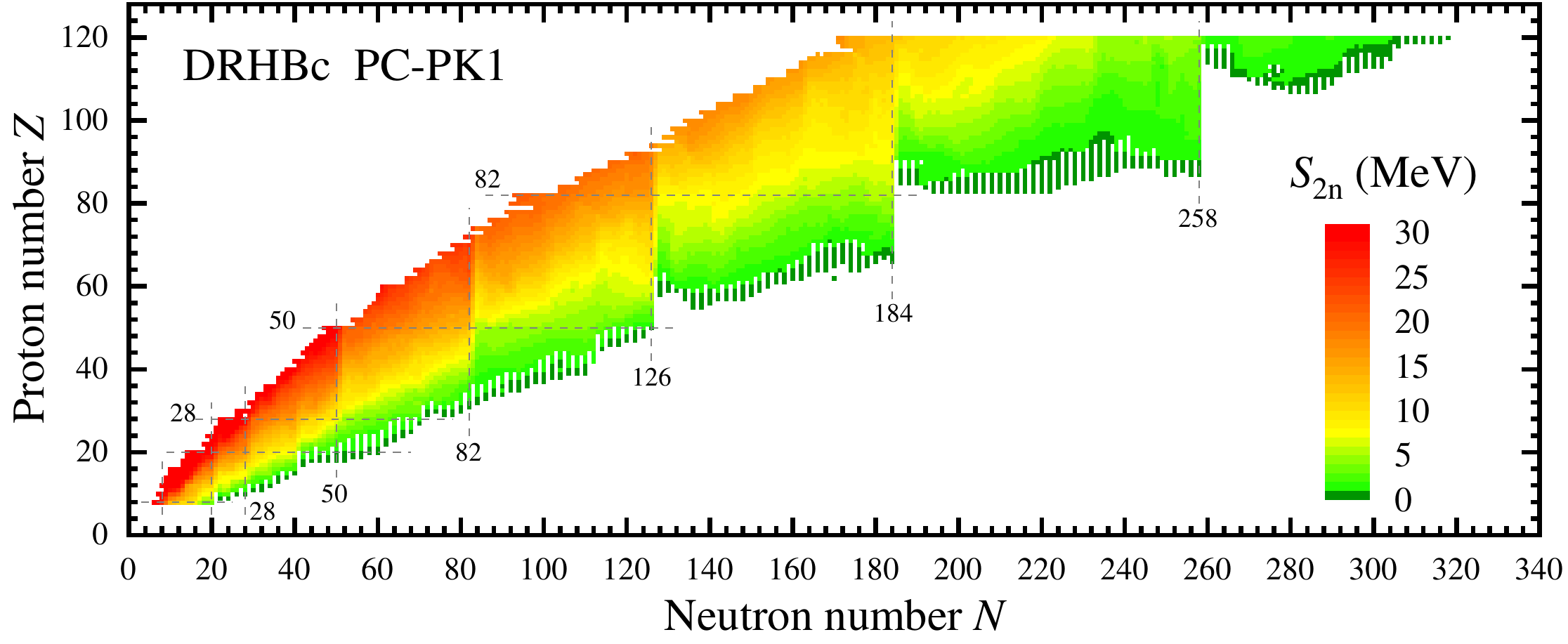}
  \caption{(Color online) Two-neutron separation energies of bound nuclei with $8 \le Z \le 120$ in the DRHBc calculations with PC-PK1 scaled by colors.}
\label{fig2}
\end{figure}

The two-neutron separation energies $S_\mathrm{2n}$ of the bound nuclei predicted by the DRHBc theory with PC-PK1 are shown in Fig.~\ref{fig2}.
For a given isotonic chain, $S_\mathrm{2n}$ increases almost monotonically with the proton number, while for a given isotopic chain, $S_\mathrm{2n}$ decreases almost monotonically with the neutron number.
The abrupt drops of $S_\mathrm{2n}$ at the $N = 20, 28, 50, 82, 126, 184$, and $258$ correspond to the known neutron magic numbers or the predicted ones~\cite{Zhang2005NPA,Li2014PLB}.
% Statistically, there are 121 nuclei with $S_\mathrm{2n} \geq 30$ MeV, 553 ones with $ 20\leq S_\mathrm{2n} < 30$ MeV, 2440 ones with $ 10\leq S_\mathrm{2n} < 20$ MeV, and 6381 ones with $ 0\leq S_\mathrm{2n} < 10$ MeV. 
% In particular,
The DRHBc calculations predict 576 weakly bound nuclei with $ S_\mathrm{2n} < 1$ MeV.
Among them, there are 253 even–even, 272 odd-$Z$ even-$N$, 25 even-$Z$ odd-$N$, and 26 odd-odd nuclei.
% that are extremely neutron-rich and lie even beyond the neutron drip lines predicted by other models. 
% In contrast, there are only 25 even-$Z$ odd-$N$ and 25 odd-odd nuclei with $ S_\mathrm{2n} < 1$ MeV, and 
For 25 even-$Z$ odd-$N$ and 26 odd-odd nuclei with $ S_\mathrm{2n} < 1$ MeV, 33 of them are located in the superheavy mass region with $104 \leq Z \leq 120$.
%统计偶偶、偶奇、奇偶、奇奇在S2n<1的个数
For nuclei with $S_\mathrm{2n} < 1 $ MeV, halo or giant halo may exist~\cite{Meng1998PRL,Meng2002PRC,Zhang2003SCG,Terasaki2006PRC}.
The research in this direction would be interesting and necessary.
% Such small two-neutron separation energies might be possible signals for two-neutron halos or giant halos, which need detailed analysis of neutron radii, density distributions and single-particle levels to confirm~\cite{Meng1998PRL,Meng2002PRC,Zhang2003SCG,Terasaki2006PRC}. The search for halo candidates based on the DRHBc mass table would be a promising future direction~\cite{Zhang2024NPR,Lu2026PRC,Sun2024NPR,Zhang2025CSB}.

%--------------------------------------------------------------------------,--------------,
% \subsubsection{Two-proton separation energies},,
\begin{figure}[htbp]
  \centering 
  \includegraphics[width=1.0\textwidth]{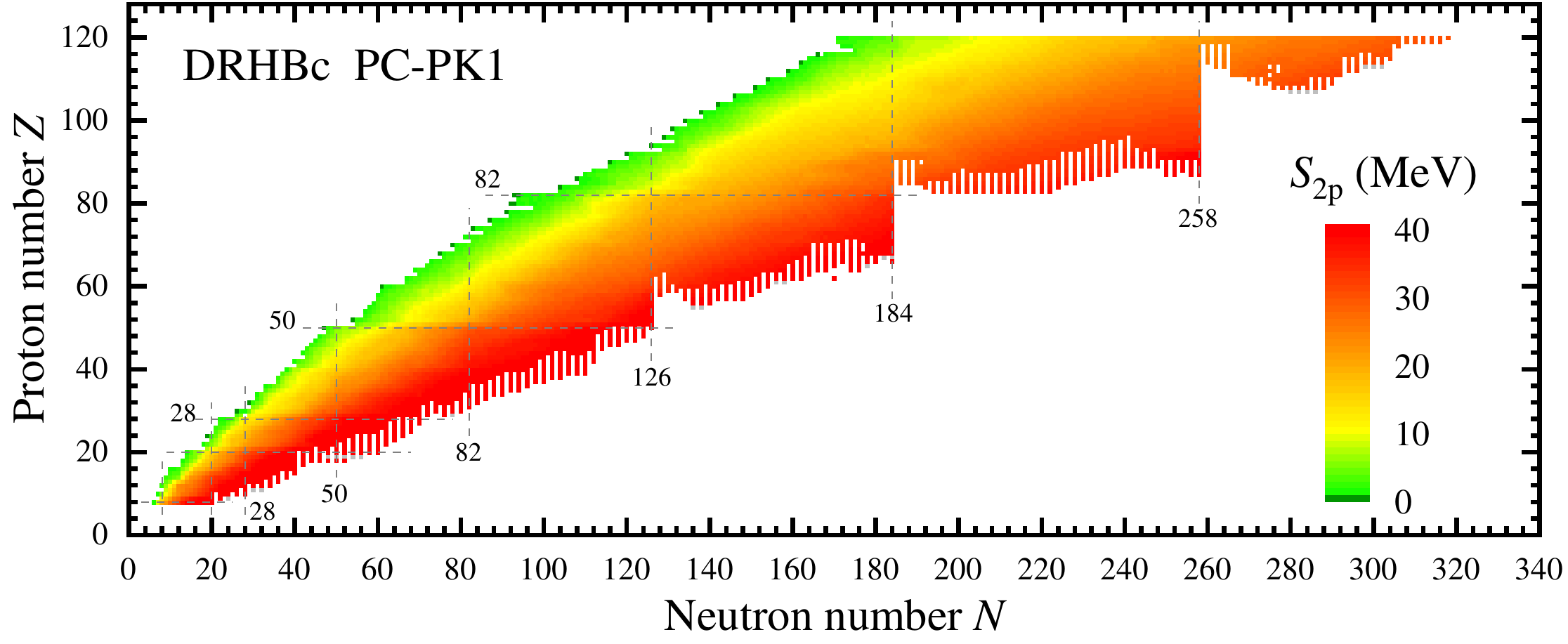}
  \caption{(Color online) The same as Fig.~\ref{fig2} but for two-proton separation energies.}
\label{fig3}
\end{figure}
%----------------------------------------------------------------------------------------

Figure~\ref{fig3} depicts the two-proton separation energies $S_\mathrm{2p}$ of the bound nuclei predicted by the DRHBc theory with PC-PK1 scaled by colors.
For a given isotonic chain, $S_\mathrm{2p}$ decreases almost monotonically with the proton number, while for a given isotopic chain, $S_\mathrm{2p}$ increases almost monotonically with the neutron number. 
The abrupt drops of $S_\mathrm{2p}$ at the $Z = 20, 28, 50$, and $82$ correspond to the proton magic numbers.
There are only 31 nuclei with $ S_\mathrm{2p} < 1$ MeV, which are all even-$Z$ nuclei. This number is much smaller than 576 for the nuclei with $ S_\mathrm{2n} < 1$ MeV. 
This is because the continuum effects are more strongly suppressed on the proton-rich side than on the neutron-rich side owing to the Coulomb barrier~\cite{Zhang2022mass,Guo2024mass}.
% The abrupt drops at proton magic numbers $20, 28, 50$, and $82$ demonstrate that the DRHBc theory reproduces the proton shell closures well.

% There are 631 nuclei with $S_\mathrm{2p} \geq 40$ MeV, 2131 ones with $ 30\leq S_\mathrm{2p} < 40$ MeV, 2915 ones with $ 20\leq S_\mathrm{2p} < 30$ MeV, 2255 ones with $ 10\leq S_\mathrm{2p} < 20$ MeV, and 1540 ones with $0 \leq S_\mathrm{2p} < 10$ MeV. 

% This is consistent with the fact that fewer halo nuclei are observed on the proton-rich side than on the neutron-rich side, due to the Coulomb barrier.

%----------------------------------------------------------------------------------------
% \subsubsection{One-neutron separation energies} 
\begin{figure}[htbp]
  \centering
  \includegraphics[width=1.0\textwidth]{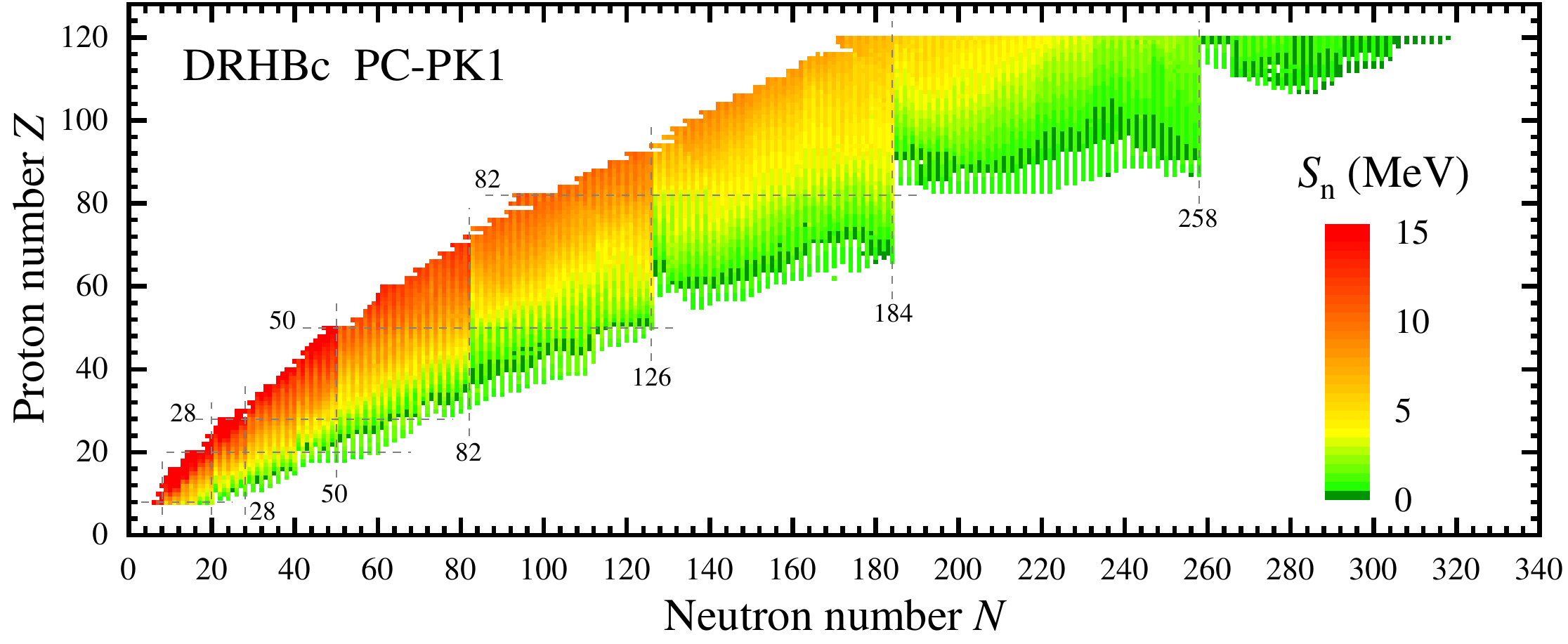}
  \caption{(Color online) The same as Fig.~\ref{fig2} but for one-neutron separation energies.}
\label{fig4}
\end{figure}
%----------------------------------------------------------------------------------------

The one-neutron separation energies $S_\mathrm{n}$ of the bound nuclei predicted by the DRHBc theory with PC-PK1 are shown in Fig.~\ref{fig4}.
Similar to $S_\mathrm{2n}$, $S_\mathrm{n}$ generally increases with the proton number for a given isotonic chain.
For a given isotopic chain, $S_\mathrm{n}$ generally decreases with the neutron number, while significant odd-even staggering occurs due to the blocking effects.
The abrupt drops of $S_\mathrm{n}$ at the $N = 20, 28, 50, 82, 126, 184$, and $258$ correspond to the known magic numbers or predicted ones.
There are in total 1430 weakly bound nuclei with $ S_\mathrm{n} < 1$ MeV and 580 nuclei with $ S_\mathrm{n} < 0.5$ MeV, of which 539 nuclei have an odd neutron number and 41 nuclei have an even neutron number.
It is noted that 36 of these 41 even-$N$ nuclei are located in the superheavy mass region with $104 \leq Z \leq 120$.
For weakly bound nuclei with $S_\mathrm{n} < 0.5 $ or 1 MeV, it is interesting to investigate whether they are candidates for halo nuclei.

%----------------------------------------------------------------------------------------
% \subsubsection{One-proton separation energies} 
\begin{figure}[htbp]
  \centering
  \includegraphics[width=1.0\textwidth]{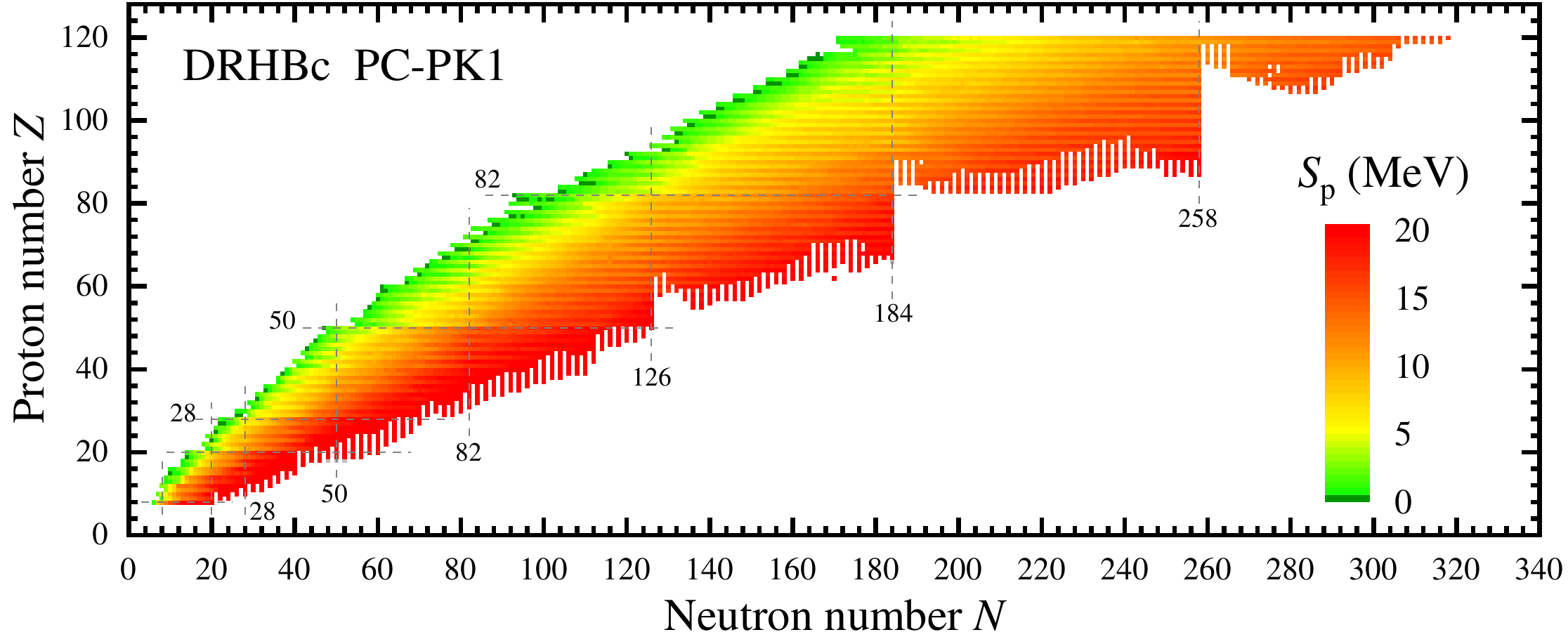}
  \caption{(Color online) The same as Fig.~\ref{fig2} but for one-proton separation energies.}
\label{fig5}
\end{figure}
%----------------------------------------------------------------------------------------

The one-proton separation energies $S_\mathrm{p}$ of the bound nuclei predicted by the DRHBc theory with PC-PK1 are shown in Fig.~\ref{fig5}. Similar to $S_\mathrm{2p}$, $S_\mathrm{p}$ generally increases with the neutron number for a given isotopic chain.
For a given isotonic chain, $S_\mathrm{p}$ generally decreases with the proton number, while significant odd-even staggering exists.
The abrupt drops of $S_\mathrm{p}$ at the $Z = 20, 28, 50$, and $82$ correspond to the proton magic numbers.
There are 211 weakly bound nuclei with $ S_\mathrm{p} < 1$ MeV.
For the 86 nuclei with $ S_\mathrm{p} < 0.5$ MeV, there are 81 odd-$Z$ nuclei and 5 even-$Z$ nuclei.
For weakly bound nuclei with $S_\mathrm{p} < 0.5 $ or 1 MeV, it is interesting to investigate whether they are candidates for halo nuclei and the competition between quantumn tunneling and Coulomb barrier in nuclei with large $Z$.

\subsection{Comparison with other predictions}\label{secC}
\subsubsection{Masses and separation energies}\label{sub:comparison_with_other_predictions}
\begin{table}\label{tab:sig}
\centering
\caption{The rms deviations for the binding energies, two-neutron separation energies, two-proton separation energies,  one-neutron separation energies, and one-proton separation energies in the DRHBc calculations with PC-PK1 with respect to the AME2020 data~\cite{AME2020(3)} in the unit of MeV, which is bolded in the chart. The results of other relativistic and non-relativistic density functional calculations are also listed for comparison. The number of the bound nuclei with available experimental data used in the calculation is listed in the last second column.}
\scalebox{0.88}{
\begin{tabular}{ccccccccc}
\hline
Model &  Density functional & ~~$\sigma(E_{\mathrm{b}})$~~ & ~~$\sigma(S_\mathrm{2n})$~~ & ~~$\sigma(S_\mathrm{2p})$~~ & ~~$\sigma(S_\mathrm{n})$~~ & ~~$\sigma(S_\mathrm{p})$~~ & Data number & Reference \\
\hline
DRHBc$^{\mathrm{w/o}~E_{\mathrm{rot}}}$ &  PC-PK1 & \textbf{2.434} & \textbf{0.920} & \textbf{0.891} & \textbf{0.669} & \textbf{0.686} & 2380 & This work \\
DRHBc$^{\mathrm{w/} ~~E_{\mathrm{rot}}}$ &  PC-PK1 & \textbf{1.444} & \textbf{0.939} & \textbf{0.998} & \textbf{0.700} & \textbf{0.777} & 2380 & This work \\
RCHB &  PC-PK1 & 8.143 & 1.477 & 1.527 & 0.865 & 0.886 & 2347 &~\cite{Xia2018ADNDT} \\
RMF+BCS & TMA & 2.040 & 0.861 & 1.071 & 0.644 & 0.802 & 2381 &~\cite{Geng2005PTP}\\
\hline
HFB &  SkM$^*$ & 6.894 & 1.092 & 1.736 & 0.682 & 0.979 & 2354 &~\cite{Erler2012Nature} \\
HFB &  SLy4 & 5.172 & 0.875 & 0.774 & 0.634 & 0.609 & 2382 &~\cite{Erler2012Nature}\\
HFB &  SV-min & 3.324 & 0.689 & 0.719 & 0.520 & 0.538 & 2382 &~\cite{Erler2012Nature} \\
HFB &  UNEDF1 & 1.983 & 0.672 & 0.751 & 0.492 & 0.543 & 2387 &~\cite{Erler2012Nature}\\
\hline
\end{tabular}

}
\label{tab2}
\end{table}

In Table~\ref{tab2}, the rms deviations of binding energies $\sigma(E_{\mathrm{b}})$, two-neutron separation energies $\sigma(S_\mathrm{2n})$, two-proton separation energies $\sigma(S_\mathrm{2p})$, one-neutron separation energies $\sigma(S_\mathrm{n})$, and one-proton separation energies $\sigma(S_\mathrm{p})$ from the present calculations with respect to the data available from AME2020~\cite{AME2020(3)} are listed. For comparison, results from several previous density functional calculations including relativistic continuum Hartree-Bogoliubov (RCHB) calculations with PC-PK1, relativistic mean field (RMF) plus BCS calculations with TMA, and Hartree-Fock-Bogoliubov (HFB) calculations with SkM$^*$, SLy4, SV-min, and UNEDF1, are also presented in the same table.
As the number of bound nuclei may vary across models, the numbers of data points used to extract the rms deviations are also listed in Table~\ref{tab2}.
For the binding energies, the rms deviations without and with rotational correction energies $E_{\mathrm{rot}}$ are 2.434 MeV and 1.444 MeV, respectively.
These deviations are simliar to those only for even-$Z$ nuclei in Ref.~\cite{Guo2024mass}, after including the data of odd-$Z$ nuclei. 
From Table~\ref{tab2}, it is shown that the DRHBc calculations with functional PC-PK1 including rotational correction provide the best description for nuclear mass.
This conclusion still holds, even if 2331 commonly bound nuclei in all the calculations are used in the evaluation. 
This success demonstrates the importance of the deformation degree of freedom and the rotational correction, as shown in the first, second, and third rows in Table~\ref{tab2}.

The separation energies in Table~\ref{tab2} have rms deviations in the range of 0.492~MeV \text{--} 1.736~MeV in comparison with the rms deviations in the binding energies ranging from 1.444~MeV to 8.143~MeV.
This reflects that the separation energies, as the derivatives of the binding energies, are reasonably described in all the nuclear models.
In comparison with the binding energies, the rotational corrections do not improve the DRHBc description for the nucleon separation energies.
The current rotational correction is taken into account through the cranking approximation.
This treatment may be not suitable for spherical and weakly deformed nuclei~\cite{Guo2024mass}.
It might be a good choice to adopt collective Hamiltonian method for treating the rotational correction. Research along this line can be found in Refs.~\cite{Sun2022CPC,ZhangXY2023PRC}.

\subsubsection{Limits of the nuclear landscape}
\label{sec:lim}
\begin{figure}[htbp]
  \centering
  \includegraphics[width=1.0\textwidth]{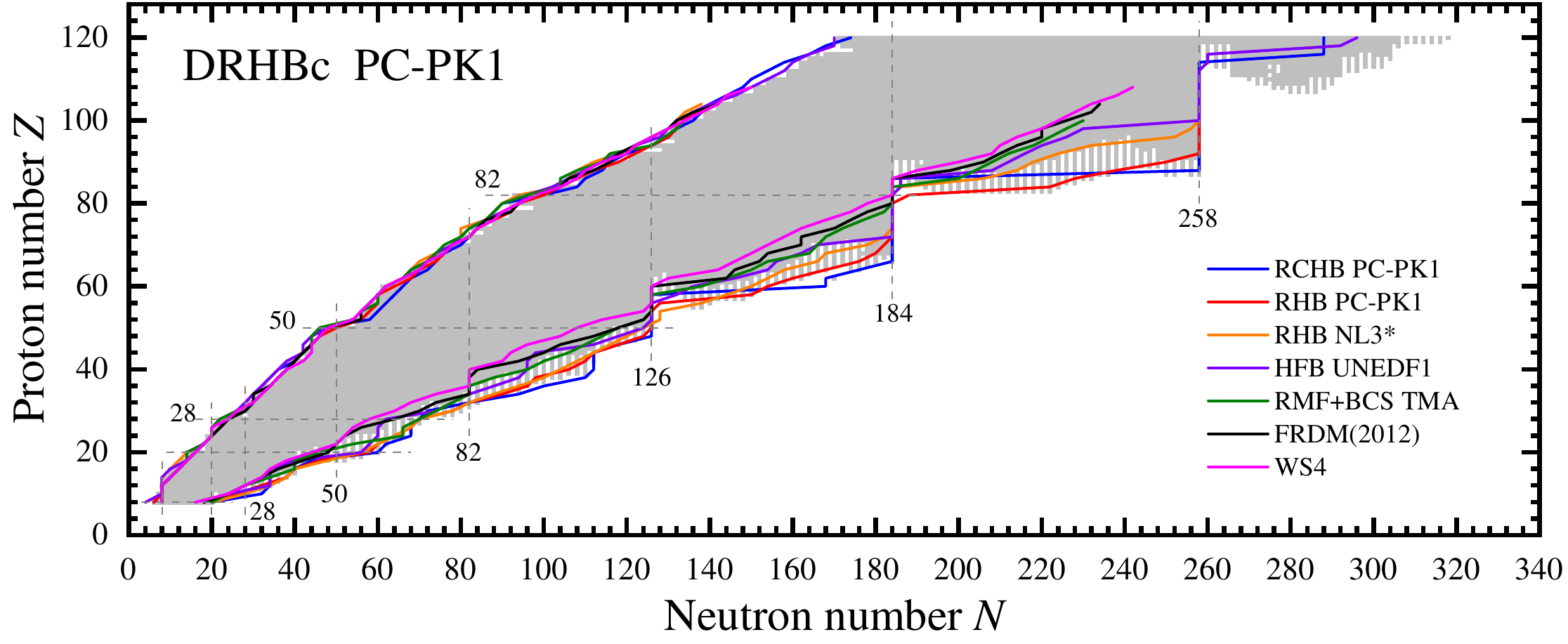}
  \caption{(Color online) Nuclear landscape from O ($Z = 8$) to $Z = 120$ predicted by the DRHBc theory with PC-PK1, consisting of 9495 bound nuclei. The nucleon drip lines predicted by different mass tables, including RCHB with PC-PK1~\cite{Xia2018ADNDT}, RHB with PC-PK1~\cite{Yang2021PRC} and with NL3*~\cite{Agbemava2014PRC}, RMF+BCS with TMA~\cite{Geng2005PTP}, HFB with UNEDF1~\cite{Erler2012Nature}, FRDM(2012)~\cite{Moller2016ADNDT}, and WS4~\cite{Wang2014PLB}, are plotted for comparison.}
\label{fig6}
\end{figure}

In Fig.~\ref{fig6}, the boundaries of the nuclear landscape predicted by the DRHBc mass table are compared with the mass models RCHB with PC-PK1~\cite{Xia2018ADNDT}, RHB with PC-PK1~\cite{Yang2021PRC} and with NL3*~\cite{Agbemava2014PRC}, RMF+BCS with TMA~\cite{Geng2005PTP}, HFB with UNEDF1~\cite{Erler2012Nature}, FRDM(2012)~\cite{Moller2016ADNDT}, and WS4~\cite{Wang2014PLB}.

As shown in Fig.~\ref{fig6}, on the proton-rich side, the proton drip lines predicted by the DRHBc mass table are close to those of other mass models in most regions.
This is because the continuum effects are suppressed due to the Coulomb barrier.
Due to the deformation effects, the proton drip lines predicted by the DRHBc mass table are more extended than those of the RCHB mass table from $Z = 52$ to 60, whereas the RCHB mass table predicts more extended proton drip lines from $Z = 106$ to 118.

On the neutron-rich side, Fig.~\ref{fig6} shows significant discrepancies in neutron drip line predicted by different theoretical models and density functionals.
The macroscopic–microscopic mass models FRDM~\cite{Moller2016ADNDT} and WS4~\cite{Wang2014PLB} predict neutron drip lines systematically closer to the valley of stability. 
The DRHBc mass table generally predicts the most extended neutron drip lines.
Due to the deformation effects, the DRHBc mass table predicts more extended neutron drip lines than the RCHB for $55 \leq Z \leq 59$, $83 \leq Z \leq 87$, and $107 \leq Z \leq 120$.
The neutron drip lines predicted by other mass models lie roughly between those of macroscopic–microscopic mass models and those of the DRHBc mass table.

On the extremely neutron-rich region, the DRHBc mass table predict the possible existence of nuclei beyond the drip lines defined by one-neutron and two-neutron separation energies.
% stable against neutron emission beyond the region of unbound nuclei against two- or multineutron emission.
Typical examples are the bound nuclei beyond $N= 258$ with $107 \le Z \le 113$, forming a peninsula of stability extending from the mainland~\cite{Zhang2021PRC,He2024PRC,He2021CPC}.
In Fig.~\ref{fig6}, some other smaller peninsulas can be seen consisting of even-$N$ nuclei $^{191,193}$Cs, $^{192, 194, 196}$Ba, $^{191, 193, ..., 205}$La, $^{192, 194, ..., 208}$Ce, $^{232}$Sm, $^{238,240}$Gd, $^{250}$Dy, $^{277, 279, ..., 305}$Bi, $^{276, 278, ..., 308}$Po, $^{337,339,343,345}$Fr, and $^{338, 340, ..., 346}$Ra.
It should be noted that the nucleus $^{232}$Sm has 8 neutrons more than its last bound isotope and 3 protons less than its next bound isotone.
In Refs.~\cite{Zhang2021PRC,He2021CPC,Pan2021PRC,He2024PRC}, the possible existence of these stability peninsulas and the underlying microscopic mechanism have been studied. It was found that the deformations, which significantly affect the single-neutron spectrum near the Fermi surfaces, play a vital role in the formation of the peninsulas of stability.
As higher-order deformations are taken into account, more nuclei that are traditionally unbound against one- and two-neutron emission may exist.

%----------------------------------------------------------------------------------------
\subsection{Pairing gaps}
% \subsection{Odd-even mass differences}

The total binding energy of an odd-$N$($Z$) nucleus has been found to be smaller than the arithmetic mean of the binding energies of the two neighboring even-$N$($Z$) nuclei, which is known as the odd-even effect~\cite{NMBP1980}. The odd-even effect reflects the pairing correlations, described by the pairing gaps which can be quantified by the three-point formula,
% The odd-even effect is an evidence of the pairing correlations and can be quantified by the corresponding pairing gaps, defined by the three-point formula,
\begin{gather}
 \Delta^{(3)}_\mathrm{n}(Z,N) = \frac{(-1)^N}{2} [2E_{\mathrm{b}}(Z,N)-E_{\mathrm{b}}(Z,N-1)-E_{\mathrm{b}}(Z,N+1)],\label{delta3n}\\
 \Delta^{(3)}_\mathrm{p}(Z,N) = \frac{(-1)^Z}{2} [2E_{\mathrm{b}}(Z,N)-E_{\mathrm{b}}(Z-1,N)-E_{\mathrm{b}}(Z+1,N)].\label{delta3p}
\end{gather}
%----------------------------------------------------------------------------------------
\begin{figure}[htbp]
  \centering
  \includegraphics[width=1.0\textwidth]{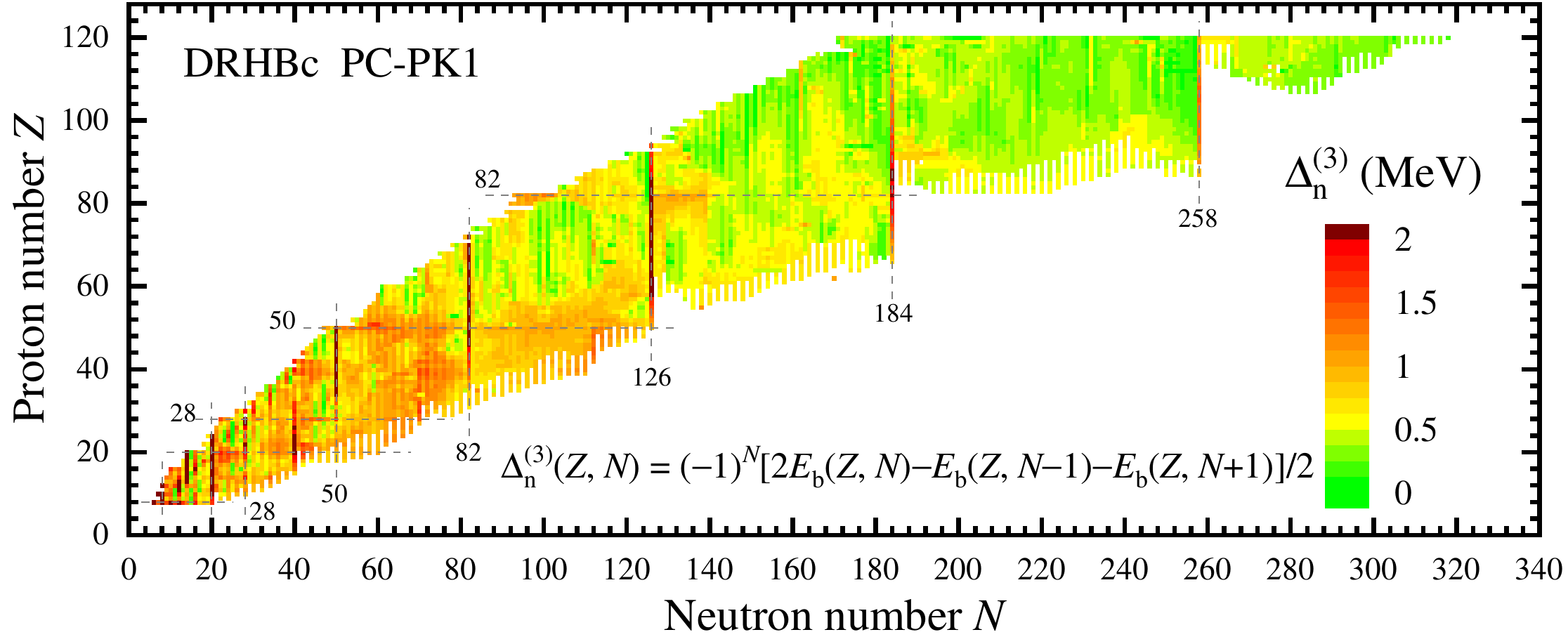}
  \caption{(Color online) Neutron pairing gaps estimated by Eq.~\eqref{delta3n} of bound nuclei with $8 \le Z \le 120$ in the DRHBc calculations with PC-PK1 scaled by colors.}
\label{fig7}
\end{figure}
%----------------------------------------------------------------------------------------

Figure~\ref{fig7} shows the neutron pairing gaps estimated by three-point formula $\Delta^{(3)}_\mathrm{n}(N,Z)$ in Eq.~\eqref{delta3n} of the bound nuclei predicted by the DRHBc theory with PC-PK1. 
% In general, the nuclei with $Z \leq 50$ have larger $\Delta^{(3)}_\mathrm{n}$ than those with $Z > 50$.
The neutron pairing gaps generally decrease with the increasing mass number $A$, which is consistent with the empirical formula $\Delta \approx 12/\sqrt{A}$~MeV~\cite{BAM1975I}.
Pronounced neutron pairing gaps appear at the traditional neutron magic numbers $N=20, 28, 50, 82$, and 126 and the predicted ones 184 and 258, owing to shell closures as discussed in detail in Sec.~\ref{tng}.
Similar situations also appear at the subshells, \textit{e.g.}, $N = 14$ and 40.
From the available experimental nuclear masses~\cite{AME2020(3)}, 2186 data for $\Delta_\mathrm{n}^{(3)}$ can be extracted and compared with experimental data.
% , and the rms deviation for $\Delta_\mathrm{n}^{(3)}$ is 0.485 MeV in DRHBc calculations.
In general, the DRHBc calculations reasonably reproduce the experimental neutron pairing gaps with the rms deviation $\sigma = 0.485$~MeV.
With the rotational correction included for the binding energies, the rms deviations for $\Delta_\mathrm{n}^{(3)}$ is 0.521~MeV.
This demonstrates that the current rotational correction does not improve the descriptions for the pairing gaps, as in the case of nucleon separation energies discussed in Sec.~\ref{sub:comparison_with_other_predictions}. 
Future better treatment for the rotational correction can improve the agreement.
% Both cases suffer from the same limitation of the cranking approximation.

%----------------------------------------------------------------------------------------
\begin{figure}[htbp]
  \centering
  \includegraphics[width=1.0\textwidth]{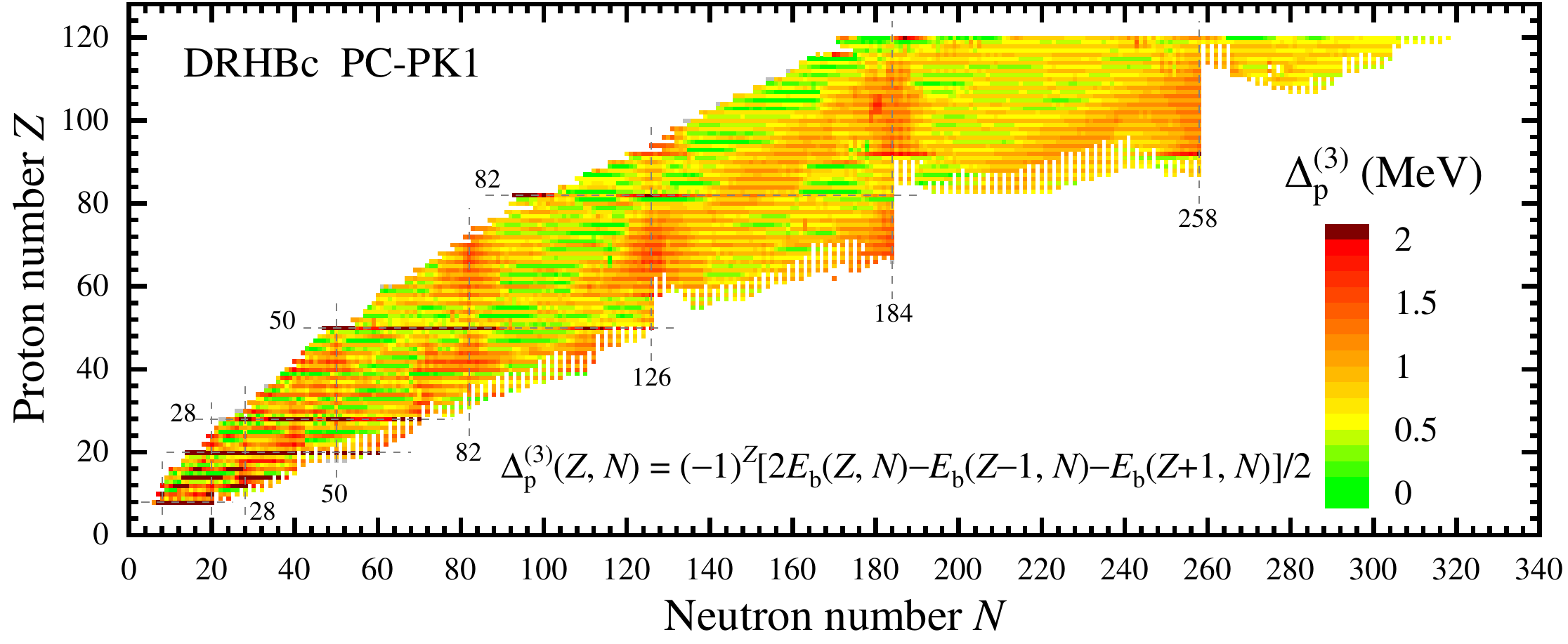}
  \caption{(Color online) The same as Fig.~\ref{fig7} but for proton pairing gaps estimated by Eq.~\eqref{delta3p}.}
\label{fig8}
\end{figure}
%----------------------------------------------------------------------------------------

Figure~\ref{fig8} shows the proton pairing gaps estimated by three-point formula $\Delta^{(3)}_\mathrm{p}(N,Z)$ in Eq.~\eqref{delta3p} of the bound nuclei.
Pronounced proton pairing gaps appear at the proton magic numbers $Z=8,20, 28, 50$ and $82$.
It is found that large $\Delta^{(3)}_\mathrm{p}$ appear at the proton mid-shell nuclei around the neutron magic numbers. Similar situations also appears in $\Delta^{(3)}_\mathrm{n}$ as shown in Fig.~\ref{fig7}.
From the available experimental nuclear masses~\cite{AME2020(3)}, 2085 data for $\Delta_\mathrm{p}^{(3)}$ can be extracted and compared with experimental data.
In general, the DRHBc mass table reasonably reproduces the experimental proton pairing gaps with $\sigma = 0.531$~MeV. 
With the rotational correction included for the binding energies, the rms deviations for $\Delta_\mathrm{p}^{(3)}$ is 0.596~MeV.

%----------------------------------------------------------------------------------------
\subsection{Pairing energies}\label{sub:pairing_energies}
% \subsubsection{Neutron pairing energies}
\begin{figure*}[htbp]
  \centering
  \includegraphics[width=1.0\textwidth]{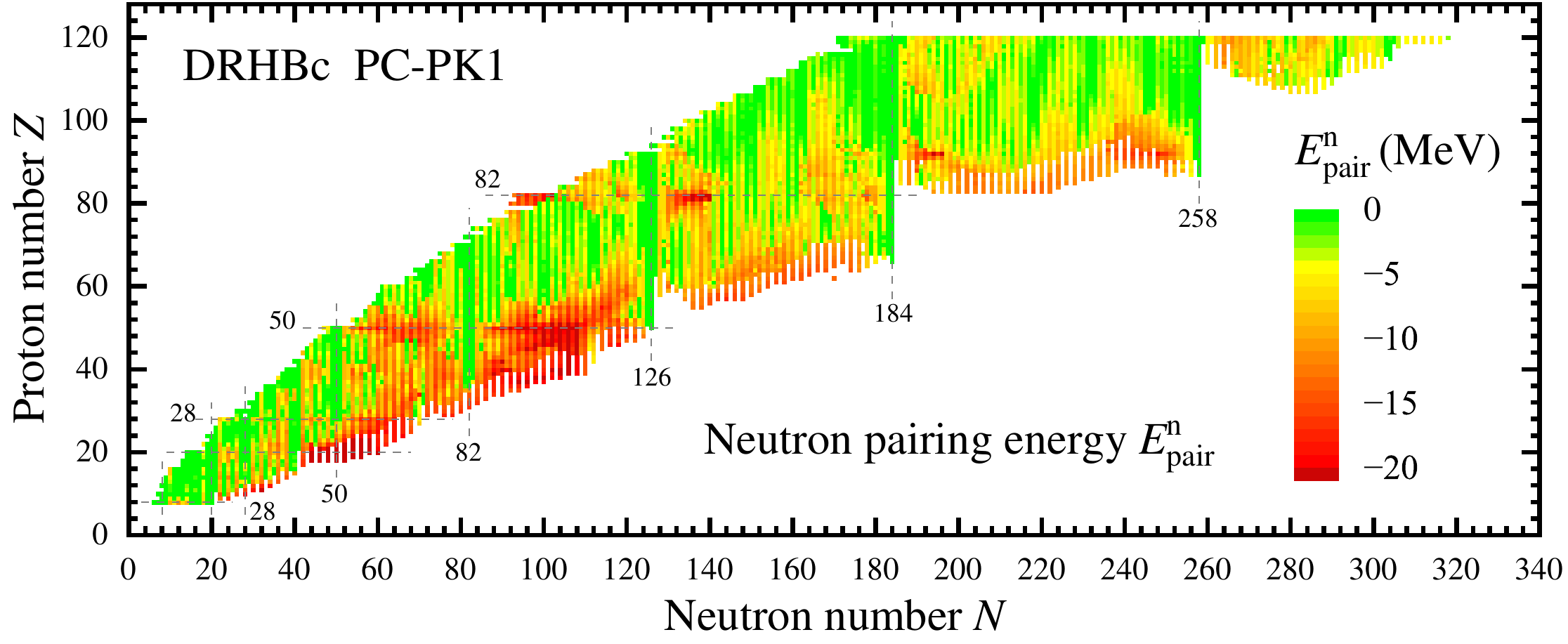}
  \caption{(Color online) Neutron pairing energies of bound nuclei with $8 \le Z \le 120$ in the DRHBc calculations with PC-PK1 scaled by colors.}
\label{fig9}
\end{figure*}
%----------------------------------------------------------------------------------------

Besides the pairing gap, the pairing energy $E_{\mathrm{pair}}$ serves as another measure of the size of pairing correlations~\cite{Afanasjev2015PRC}. With a zero-range pairing force adopted, the pairing field $\Delta(\bm{r})$ is local, and the pairing energy is calculated as,
\begin{gather}
 E_\mathrm{pair}= - \frac{1}{2} \int d^3 \bm{r} \kappa(\bm{r})\Delta(\bm{r}),
\end{gather}
where $\kappa(\bm{r})$ denotes the pairing tensor. To examine the pairing correlations from a global perspective, the pairing energies for bound nuclei with $8\leq Z \leq 120$ in DRHBc calculations have been analyzed.

Figure~\ref{fig9} shows the neutron pairing energies $E_{\mathrm{pair}}^\mathrm{n}$ of bound nuclei with $8 \le Z \le 120$.
Obvious odd-even staggering along isotopic chains can be observed in $E_{\mathrm{pair}}^\mathrm{n}$, due to the blocking effects of the unpaired neutrons.
The neutron pairing energies $E_{\mathrm{pair}}^\mathrm{n}$ vanish near the neutron magic numbers $N = 8, 20, 28, 50, 82$ and $126$ and the predicted ones $N = 184$ and $258$, which are consistent with the shell closure characteristics discussed in Sec~\ref{tns}.
For with $N<20$, the neutron pairing energies $E_{\mathrm{pair}}^\mathrm{n}$ are quite small due to the low single-neutron level density.
Pronounced neutron pairing energies $E_{\mathrm{pair}}^\mathrm{n}$ are observed for the nuclei in the middle of the neutron shell and near the neutron drip lines.
The enhanced neutron pairing energies near the neutron drip lines at the neutron magic numbers $N = 28, 50$, and 82 indicate the weakening of the corresponding traditional shell closures.

%----------------------------------------------------------------------------------------
% \subsubsection{Proton pairing energies}
\begin{figure*}[htbp]
  \centering
  \includegraphics[width=1.0\textwidth]{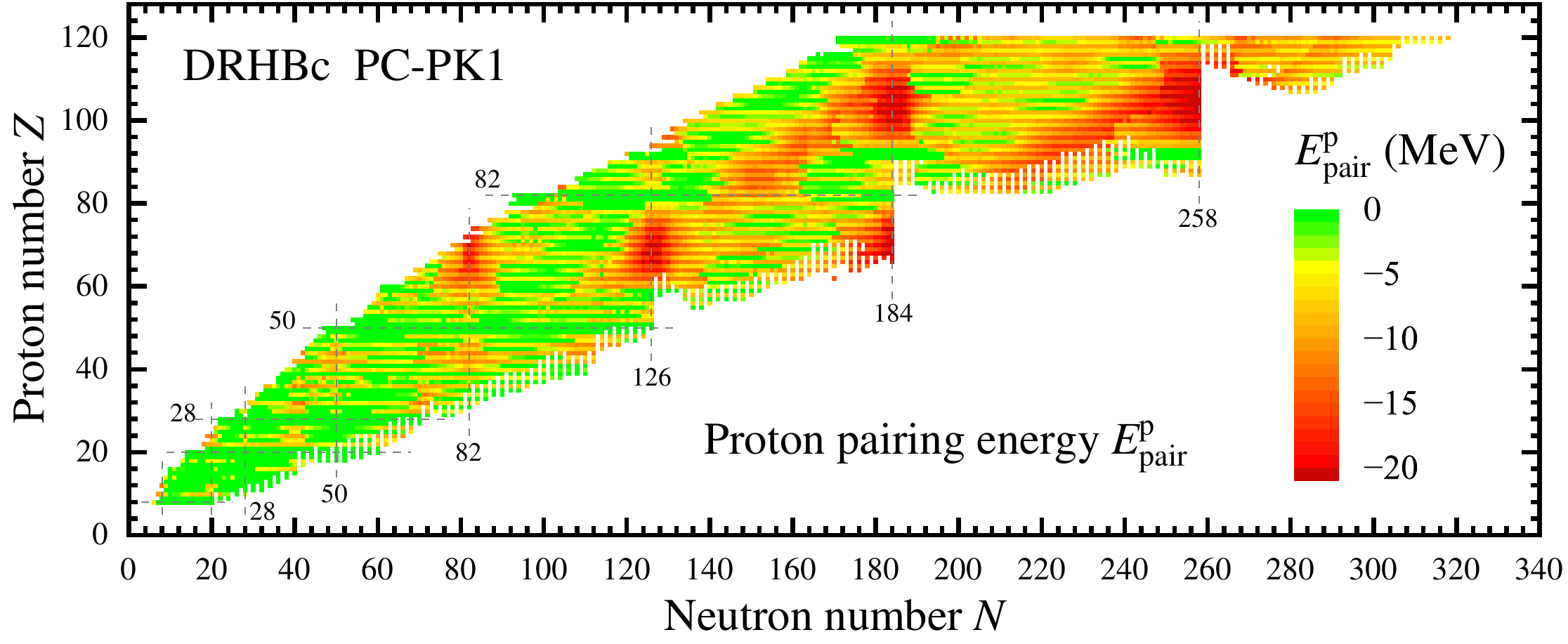}
  \caption{(Color online) The same as Fig.~\ref{fig9} but for proton pairing energies.}
\label{fig10}
\end{figure*}
%----------------------------------------------------------------------------------------

The proton pairing energies $E_{\mathrm{pair}}^\mathrm{p}$ for the nuclei with $8 \le Z \le 120$ are presented in Fig.~\ref{fig10}.
% They share many features with neutron pairing energies. 
Similar as the case for neutron, the odd-even staggering in $E_{\mathrm{pair}}^\mathrm{p}$ also exists along isotonic chains.
For most nuclei, the proton pairing energies $E_{\mathrm{pair}}^\mathrm{p}$ vanish near the proton magic numbers $Z = 8, 20, 28, 50,$ and $82$.
It is noticed that the proton pairing energies $E_{\mathrm{pair}}^\mathrm{p}$ vanish around the nuclei with $Z = 120$ and $N= 184$, indicating that $Z = 120$ may be the next proton magic number after $Z=82$~\cite{Zhang2005NPA,Li2014PLB}.
For nuclei with $Z<20$, the proton pairing energies $E_{\mathrm{pair}}^\mathrm{p}$ are quite small due to the low single-proton level density.
For the proton mid-shell nuclei with neutron number around the magic numbers, pronounced proton pairing energies $E_{\mathrm{pair}}^\mathrm{p}$ are observed.

%----------------------------------------------------------------------------------------
\subsection{Two-nucleon gaps}\label{tng}

The two-neutron gap $\delta_\mathrm{2n}$ and the two-proton gap $\delta_\mathrm{2p}$ are defined, respectively, as
\begin{gather}
\delta_\mathrm{2n}(Z,N) = S_\mathrm{2n}(Z,N) - S_\mathrm{2n}(Z,N+2),\\
\delta_\mathrm{2p}(Z,N) = S_\mathrm{2p}(Z,N) - S_\mathrm{2p}(Z+2,N).
\end{gather}
The peaks in these two-nucleon gaps, indicating the drastic change in the two-nucleon separation energies, serve as signatures for magic numbers~\cite{Zhang2005NPA,Li2014PLB,Liu2020PLB}.
We therefore focus on the shell and subshell closures indicated by the calculated two-nucleon gaps.

%----------------------------------------------------------------------------------------
% \subsubsection{Two-neutron gaps}\label{tng}
\begin{figure*}[htbp]
  \centering
  \includegraphics[width=1.0\textwidth]{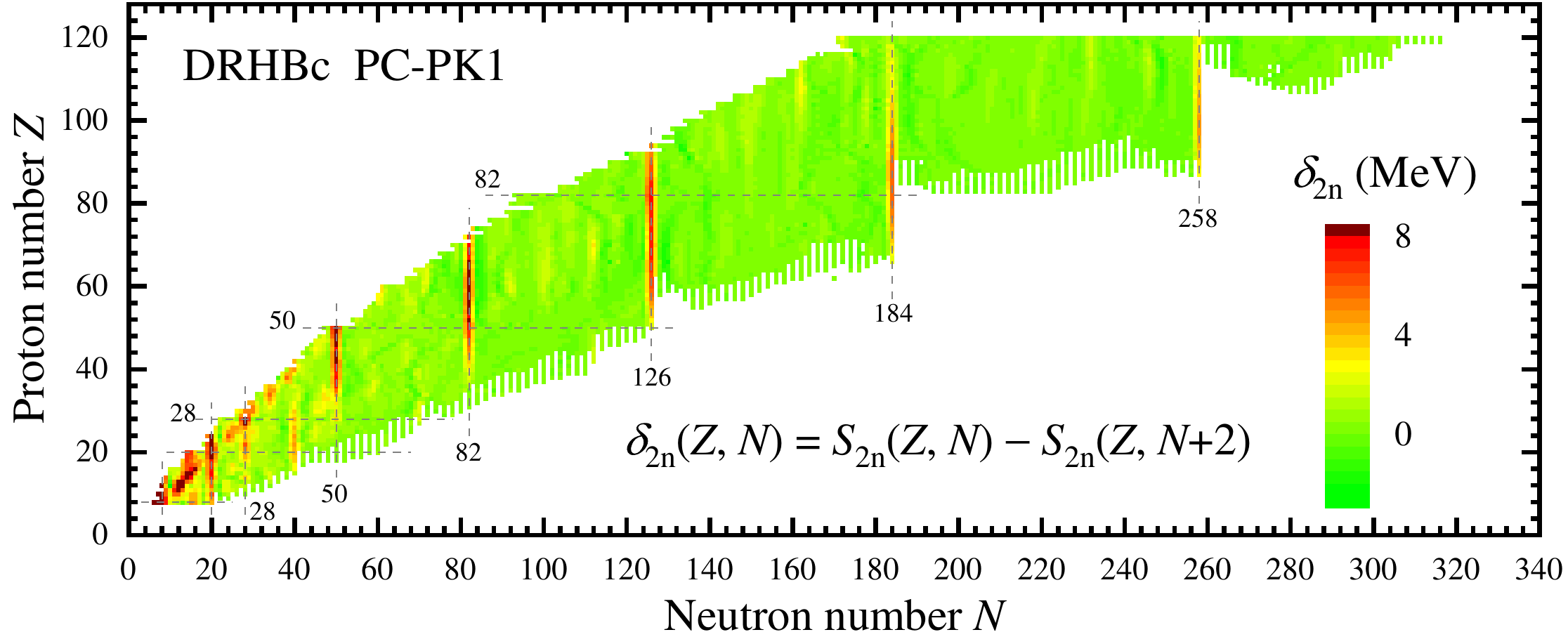}
  \caption{(Color online) Two-neutron gaps $\delta_\mathrm{2n}$ of bound nuclei with $8 \le Z \le 120$ in the DRHBc calculations with PC-PK1 scaled by colors.}
\label{fig11}
\end{figure*}
%----------------------------------------------------------------------------------------

Figure~\ref{fig11} presents the two-neutron gaps $\delta_\mathrm{2n}$ for the nuclei with $8 \le Z \le 120$ in the calculations.
Pronounced two-neutron gaps $\delta_\mathrm{2n}$ appear at the traditional neutron magic numbers $N = 8, 20, 28, 50, 82$, and 126 and the predicted ones 184 and 258.
The two-neutron gaps $\delta_\mathrm{2n}$ at neutron magic numbers become weaker or even vanish near the neutron drip lines.
This may suggest the quenching or even collapse of the traditional neutron shell closures in the neutron-rich region, which is also discussed in Sec.~\ref{sub:pairing_energies}.
The slightly pronounced two-neutron gaps $\delta_\mathrm{2n}$ appear at $N=40$ with $16 \leq Z \leq 32$ and $N=162$ with $104 \leq Z \leq 114$, due to the possible shell closure.

%----------------------------------------------------------------------------------------
% \subsubsection{Two-proton gaps}
\begin{figure*}[htbp]
  \centering  
  \includegraphics[width=1.0\textwidth]{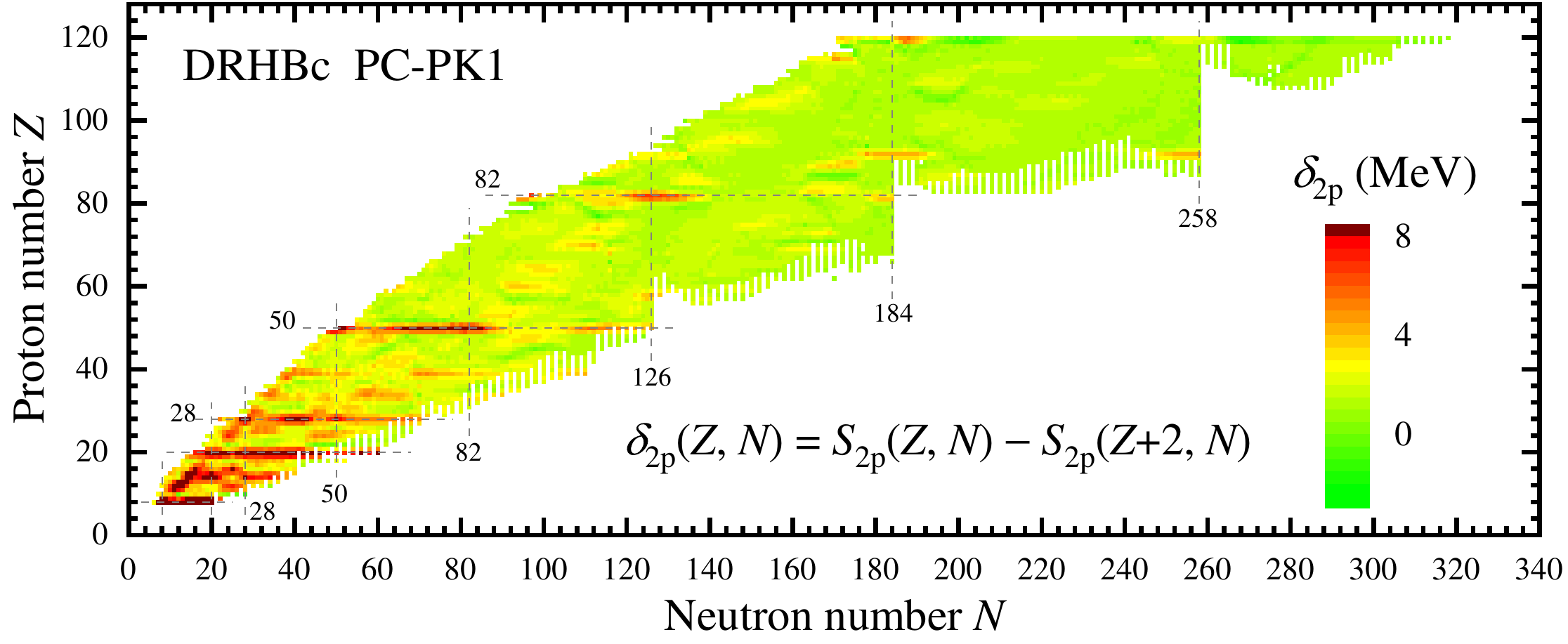}
  \caption{(Color online) The same as Fig.~\ref{fig11} but for two-proton gaps $\delta_{2p}$.}
\label{fig12}
\end{figure*}
%----------------------------------------------------------------------------------------

Figure~\ref{fig12} shows the two-proton gaps $\delta_\mathrm{2p}$ for the nuclei with $8 \le Z \le 120$.
Pronounced two-proton gaps $\delta_\mathrm{2p}$ appear at the traditional proton magic numbers $Z= 8, 20, 28, 50$, and $82$.
The slightly pronounced two-proton gaps around the nuclei with $Z = 120$ and $N = 184$ indicate that $Z = 120$ may be the next
proton magic number.
For nuclei at $Z=50$ with $90 \le N \le 100$ and $Z=82$ with $140 \le N \le 180$, the two-proton gaps $\delta_\mathrm{2p}$ are relatively small, indicating that these nuclei may deviate from the closed shell.
The pronounced two-proton gaps also appear in the nuclei around $Z=14$ and along the $N \sim Z$ band.
The slightly pronounced two-proton gaps $\delta_\mathrm{2p}$ also appear in $Z = 92$ isotopes around $N = 126, 184$, and 258. $Z = 92$ has been considered as a pseudo shell in the previous relativistic density functional calculations~\cite{Geng2006CPL}.

%----------------------------------------------------------------------------------------
\subsection{Alpha decay energies}
\begin{figure*}[htbp]
  \centering
  \includegraphics[width=1.0\textwidth]{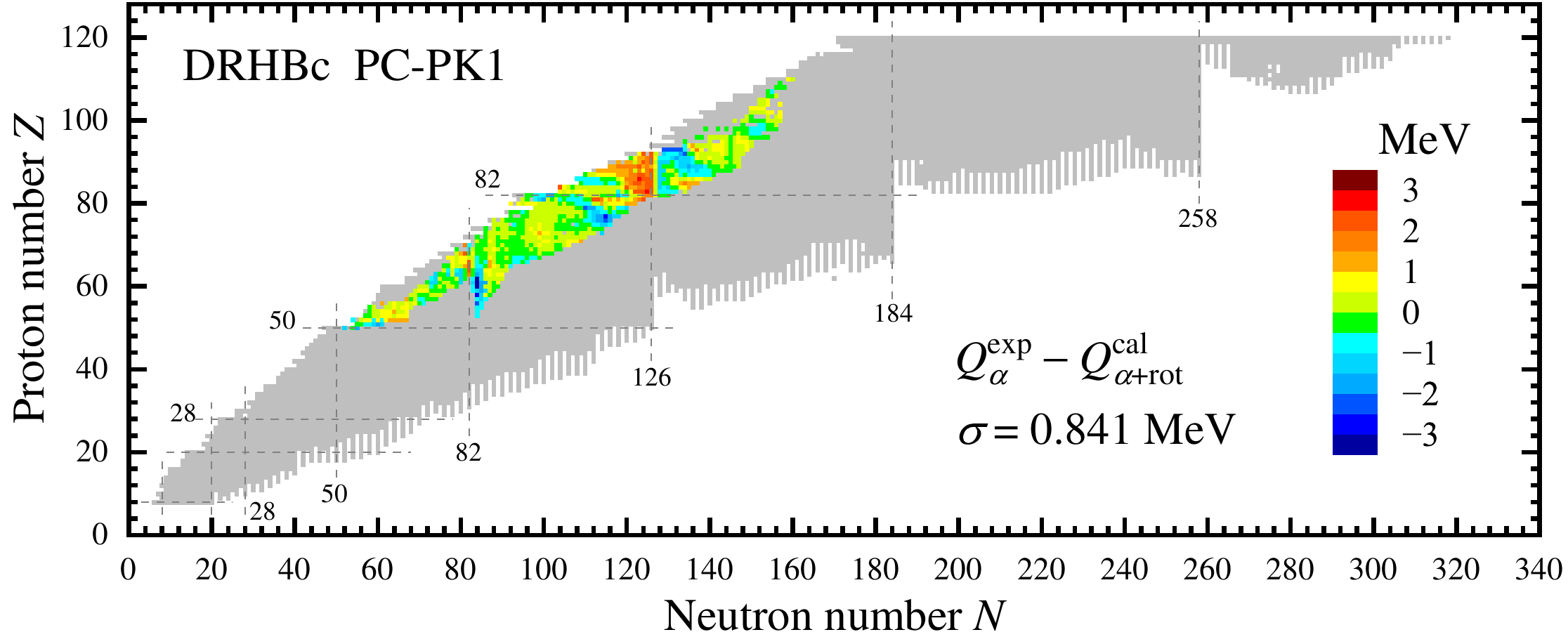}
  \caption{(Color online) The energy differences between the experimental $\alpha$ decay energy $Q_\alpha$~\cite{AME2020(3)} and the DRHBc calculations with PC-PK1 including rotational corrections, scaled by colors.}
\label{fig13}
\end{figure*}
%----------------------------------------------------------------------------------------

The $\alpha$ decay energies $Q_\alpha$ can be extracted by the binding energies,
\begin{equation}
Q_\alpha = E_{\mathrm{b}}(Z-2,N-2)+E_{\mathrm{b}}(2,2)-E_{\mathrm{b}}(Z,N),
\end{equation}
where the experimental binding energy of $^{4}$He, $E_{\mathrm{b}}(2,2)$, is adopted~\cite{AME2020(3)}. In Fig.~\ref{fig13}, the differences between the 1090 experimental data available~\cite{AME2020(3)}, $Q^\mathrm{exp}_\alpha > 0$, and the DRHBc calculations including rotational corrections $Q^\mathrm{cal}_\mathrm{\alpha+rot}$ are shown.
The rms deviation without the rotational corrections $\sigma(Q^\mathrm{cal}_\alpha)=0.948$~MeV. 
Incorporating the rotational corrections, $\sigma(Q^\mathrm{cal}_\mathrm{\alpha+rot})=0.841$~MeV. 
Compared with the RCHB calculations~\cite{Zhang2016CPC}, $\sigma(Q_\alpha)\approx2$~MeV, the description of $Q_\alpha$ has been significantly improved by including deformation in the DRHBc theory.
In general, the DRHBc mass table has achieved good agreement for $Q_\alpha$.
Several nuclei exhibit relatively large deviations in $Q_\alpha$.
For example, the DRHBc theory overestimate the $Q_\alpha$ of $^{144}$Nd, $^{145}$Pm, and $^{146}$Sm more than 3 MeV.
As shown in Fig.~\ref{fig1}, this is because the calculations underestimate their binding energies and overestimate the binding energies of their daughter nuclei.
The DRHBc theory underestimate the $Q_\alpha$ of most $N = 126$ isotones more than 2 MeV.
This is because the calculations overestimate the binding energies of most $N = 126$ isotones.
Future work is expected to improve the description of $Q_\alpha$ and to estimate systematically the $\alpha$ decay half-lives.

%----------------------------------------------------------------------------------------
\subsection{Rms radii}
\subsubsection{Charge radii}
\begin{figure*}[htbp]
  \centering
  \includegraphics[width=1.0\textwidth]{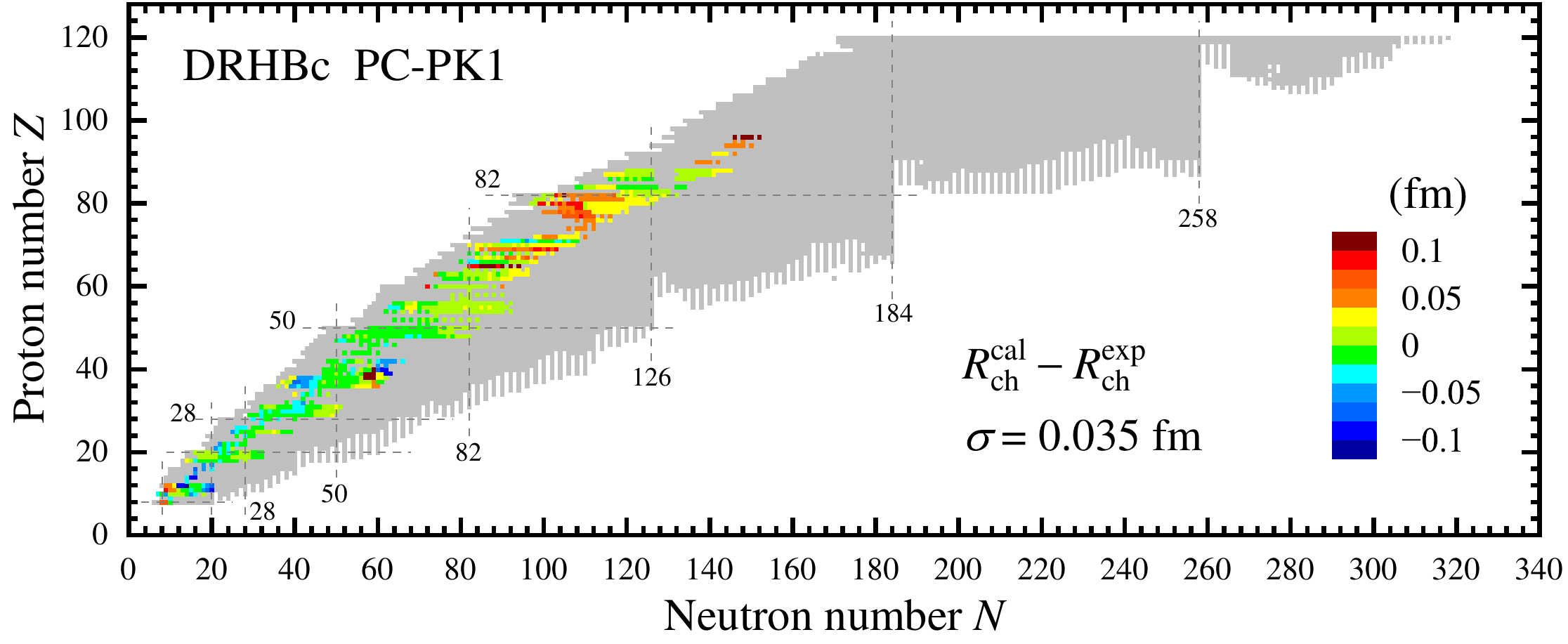}
  \caption{(Color online) The differences between the 1014 experimental charge radii~\cite{Angeli2013ADNDT,Li2021ADNDT} and the DRHBc calculations with PC-PK1, scaled by colors.
  }
\label{fig14}
\end{figure*}
%----------------------------------------------------------------------------------------

Nuclear charge radius is a pivotal observable for probing nuclear size, as it can be precisely determined via electromagnetic interactions. 

The rms radius is calculated as
\begin{equation}
R_{\tau} = \sqrt{\frac{\int d^3 \bm{r} [r^2 \rho_\tau (\bm{r})]}{N_\tau}},
\end{equation}
where $\tau$ represents neutron, proton, or the nucleon, and $\rho_\tau$ and $N_\tau$ are the corresponding vector density and particle number, respectively. The rms charge radii is simply calculated as
\begin{equation}
R_\mathrm{ch} = \sqrt{R_\mathrm{p}^2 + 0.64~\mathrm{fm}^2}.
\end{equation}
In Fig.~\ref{fig14}, the differences between the 1014 experimental charge radii $R^\mathrm{exp}_\mathrm{ch}$~\cite{Angeli2013ADNDT,Li2021ADNDT} and the DRHBc calculations $R^\mathrm{cal}_\mathrm{ch}$ are scaled by colors.
Overall, the DRHBc calculations reproduce the experimental data well, with the rms deviation $\sigma(R_\mathrm{ch})=0.035$~fm, which is slightly better than $\sigma(R_\mathrm{ch}) = 0.036$~fm in the RCHB mass table~\cite{Xia2018ADNDT}.
In Ref.~\cite{CongP2025PRC}, the charge radii and their deformation correlation for even-$Z$ nuclei in DRHBc mass table were systematically studied, and the deformation was found to play a crucial role in describing charge radii for mid-shell nuclei.

In Fig.~\ref{fig14}, relatively large discrepancies are noted for the Tb ($Z=65$), Ho ($Z=67$), and Tm ($Z=69$) isotopes, nuclei around $Z = 38$ and $N=58$, nuclei around $Z = 78$ and $N=104$, and the Cm ($Z = 96$) isotopes.
For Tb, Ho, and Tm isotopes, their deviations from the data are significantly larger than those of their even-$Z$ neighbors.
The potential energy curves (PECs) in this region exhibit two minima, indicating the possible presence of shape coexistence.
This phenomenon may be related to the large deviations of charge radii in these odd-$Z$ isotopes and also the relatively large uncertainties in the experimental charge radii of Tb isotopes~\cite{Angeli2013ADNDT}.
Further research along this line is necessary~\cite{CongP2025PRC,MunHa2024PRC,Mun2023PLB}. 
For nuclei around $Z = 38$ and $N=58$, and nuclei around $Z = 78$ and $N=104$, the DRHBc results generally overestimate the charge radii due to the overestimation of quadrupole deformation in these regions~\cite{Zhang2022mass}.
For the $N = 146$, 148, and 150 isotones, an unusual behavior is observed, \textit{i.e.}, the experimental charge radii of $^{242,244,246}$Cm are smaller than those of their respective Am ($Z = 95$), Pu ($Z = 94$), and U ($Z = 92$) isotones~\cite{Angeli2013ADNDT}.
This unusual behavior is difficult to reproduce by density functional theories~\cite{Agbemava2014PRC}.
The overestimation in charge radii of Cm isotopes is related to the unusual behavior.

%----------------------------------------------------------------------------------------
\subsubsection{Neutron rms radii}\label{nrr}
\begin{figure*}[htbp]
  \centering
  \includegraphics[width=0.7\textwidth]{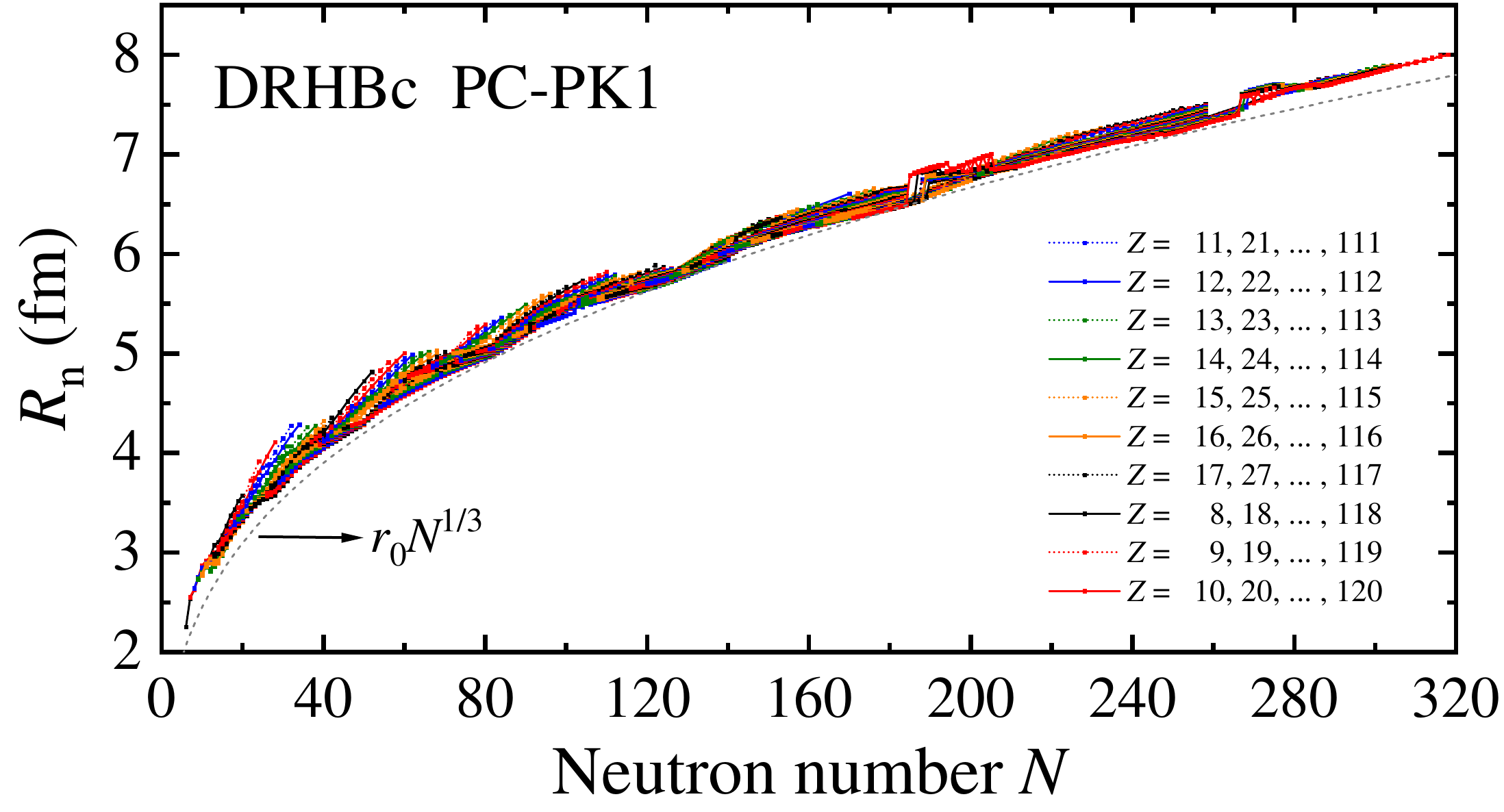}{}
  \caption{(Color online) Neutron rms radii for nuclei with $8 \le Z \le 120$ from the DRHBc calculations with PC-PK1 as a function of the neutron number. The empirical formula $r_0 N^{1/3}$, with $r_0 = 1.140$~fm determined from the calculated neutron rms radius of $^{208}$Pb, is plotted for guidance.}
\label{fig15}
\end{figure*}
%----------------------------------------------------------------------------------------

Figure~\ref{fig15} shows the neutron rms radii $R_\text{n}$ for the nuclei with $8 \le Z \le 120$ from the DRHBc calculations. The empirical formula $R_\text{n} = r_0N^{1/3}$ with $r_0 = 1.140$~fm determined from the calculated neutron rms radius of $^{208}$Pb is drawn as a dashed line for guidance.
Globally, the systematic trend of the neutron radii aligns with the empirical formula, but pronounced deviations emerge in extremely neutron-rich nuclei near the drip line, \textit{e.g.}, the near-drip-line Na, Mg, K, and Ca isotopes.
These deviations, attributed to extended neutron density distributions, can be regarded as signals for potential halos or giant halos~\cite{Meng1998PRL}.
For nuclei with $106 \leq Z \leq 120$, abrupt increases in $R_\text{n}$ are observed after the predicted shell closures at $N = 184$ and 258.

%----------------------------------------------------------------------------------------
\begin{figure*}[htbp]
  \centering
  \includegraphics[width=1.0\textwidth]{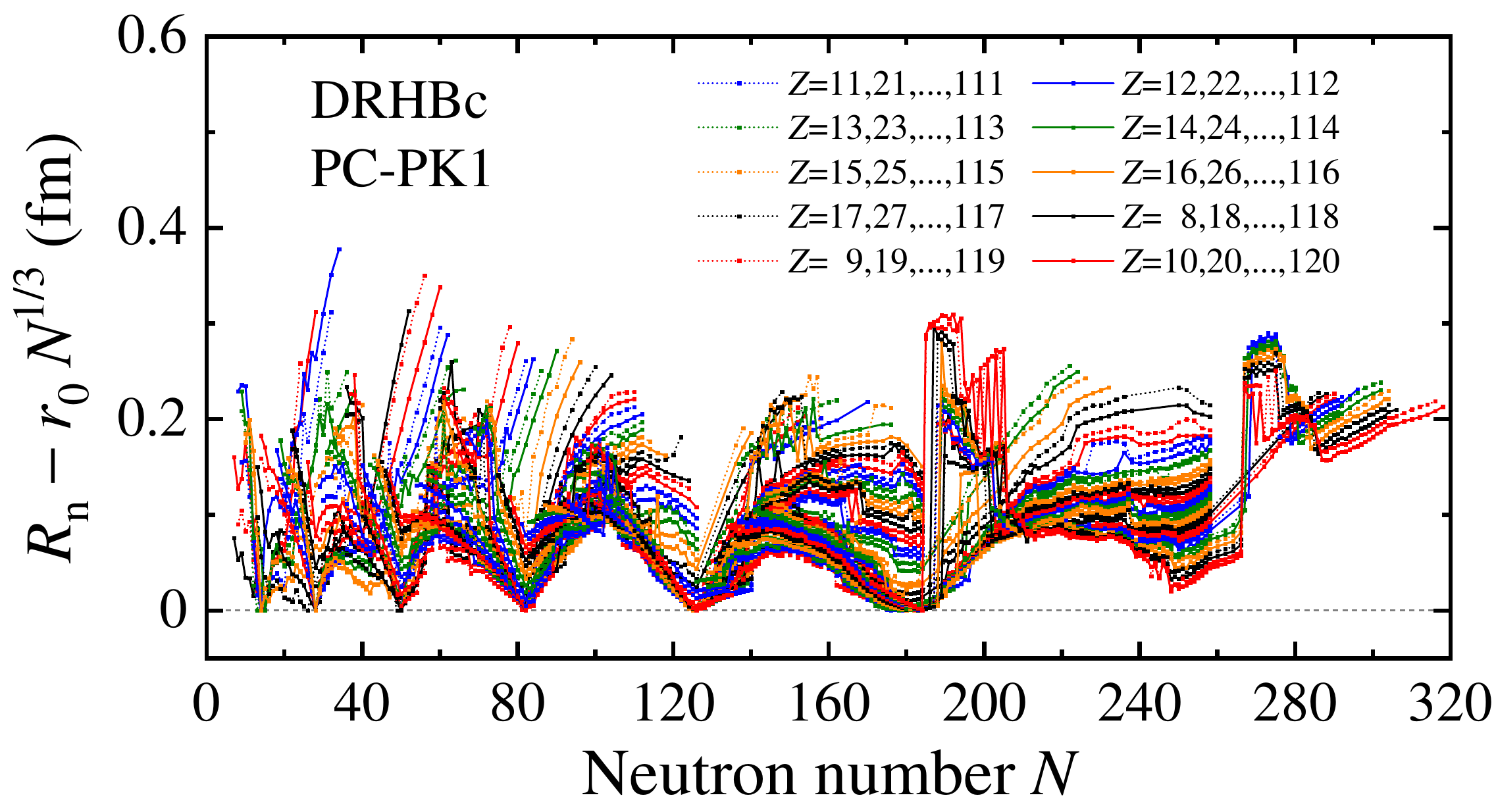}
  \caption{(Color online) Deviations of the DRHBc calculated neutron rms radii $R_n$ from the empirical formula $r_0 N^{1/3}$ for nuclei with $8 \le Z \le 120$. For each isotopic chain, $r_0$ is chosen as the smallest ratio $R_n/N^{1/3}$ to ensure the non-negative deviations.}
\label{fig16}
\end{figure*}
%----------------------------------------------------------------------------------------
 
As a detailed examination, the deviations of the calculated neutron rms radii $R_n$ from the empirical formula $r_0 N^{1/3}$ are depicted in Fig.~\ref{fig16}.
For each isotopic chain, $r_0$ is chosen as the minimum value of the ratio $R_n/N^{1/3}$.
Most minima of the deviations occur at the neutron magic numbers $N=28,50,82$, and $126$, as well as the predicted one $184$.
These isotopes typically exhibit most compact neutron distributions in their respective isotopic chains.
Minima do not occur at the neutron magic number $N = 20$ and the predicted one $258$.
The minima at $N = 14$ and the local minima at $N = 248$ indicate the possible shell closures.
It is also noted that the decrease of the deformations at $N = 14$ and $248$, which will be shown in Fig.~\ref{fig17}, may be related with the minima in these regions.
The large deviations near the neutron drip line suggest potential halo or giant halo structure, such as those in Mg~\cite{Zhou2010PRC,Li2012PRC} and Ca~\cite{Meng2002PRC} isotopes.
For nuclei with $106 \leq Z \leq 120$, the abrupt increases of the deviations are observed after the predicted shell closures at $N = 184$ and 258 in Fig.~\ref{fig16}.

%----------------------------------------------------------------------------------------
\subsection{Quadrupole deformation and potential energy curves}
\subsubsection{Quadrupole deformation}\label{qude}
\begin{figure*}[htbp]
  \centering
  \includegraphics[width=1.0\textwidth]{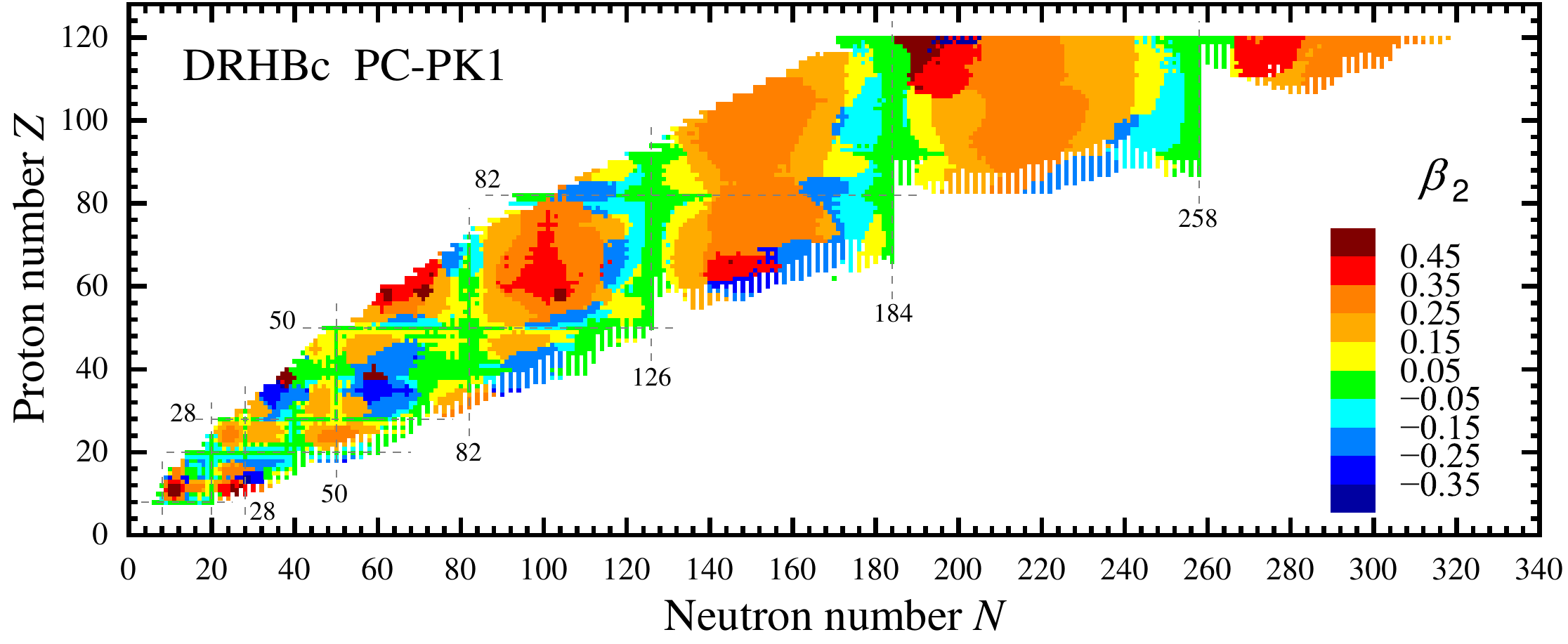}
  \caption{(Color online) Quadrupole deformations $\beta_2$ from the DRHBc calculations with PC-PK1 for the nuclei with $8 \le Z \le 120$ scaled by colors.}
\label{fig17}
\end{figure*}
%----------------------------------------------------------------------------------------

Figure~\ref{fig17} displays the quadrupole deformation parameters $\beta_2$ from the DRHBc calculations for the nuclei with $8 \le Z \le 120$.
A total of 701 nuclei with $|\beta_2|<0.001$, including 539 even-even nuclei, 64 even-odd nuclei, 87 odd-even nuclei, and 11 odd-odd nuclei, are identified as spherical ones from the DRHBc calculations.
The dominance of the even-even nuclei highlights the polarization effect of unpaired nucleons.
As expected, spherical and near-spherical nuclei often appear around neutron and proton magic numbers.
Besides, many spherical and near-spherical nuclei also appear around $Z=40$ or $N=40$, which may suggest the potential shell closure.
For $Z = 92$ isotopes around $N = 126, 184$, and 258, more spherical and near-spherical nuclei occur than those in the neighboring isotopic chains, attributed to the large shell gap at $Z = 92$.
Notably, the nuclei around $N=28,50$, and $82$ deviate from the spherical shape near neutron drip lines. 
This may suggest the quenching or collapse of the traditional shell closures in the extremely neutron-rich region, as discussed in Sec.~\ref{sub:pairing_energies} and Sec.~\ref{tng}.

Besides the 701 spherical nuclei, 6443 prolate and 2351 oblate nuclei are found, with prolate nuclei significantly outnumbering oblate ones. 
Generally, nuclear deformations gradually develop when moving away from the shell closure, typically leading to well-deformed nuclei at mid-shell. 
Generally, prolate shapes develop from the beginning of the shell, and the oblate shapes appear near the end of the shell in the DRHBc calculations.
The dominance of prolate over oblate shapes is primarily attributed to more energetically favorable orbitals on the prolate side~\cite{CASTEN2000}, which is demonstrated by the DRHBc investigations with Te, Xe, and Ba isotopes~\cite{Guo2023PRC}, Ho~\cite{ZhangJZ2026PRC}, and Bk~\cite{Huang2025PRC} chains.
The nuclei near the neutron drip lines are generally assumed to be spherical.
Interestingly, it is noted that the deformations develop at the neutron drip lines for $Z = 9~\text{--}~13$, 18, 19, $29~\text{--}~38$, $55~\text{--}~65$, $83~\text{--}~86$, and $107~\text{--}~120$.

In Fig.~\ref{fig17}, several abrupt shape changes from well-deformed prolate shape to well-deformed oblate shape are observed.
Typical examples can be found for nuclei with $60 \leq Z \leq 71$ around $N = 115$. 
Such shape changes are mainly attributed to the competition between the prolate and oblate minima with close energies, which will be discussed in detail in the following section.
For these nuclei with two or more minima with similar energies, possible shape coexistence and/or triaxial deformation is expected.
For instance, triaxiality in the light nuclear region with $Z < 20$ and in the heavier region with $114 \le N \le 120$ for $54 \le Z \le 78$ has been investigated based on relativistic density functional theory~\cite{Yao2010PRC,Yao2011PRC,Yang2021PRC,Zhang2023PRCL}.
For nuclei with $106 \leq Z \leq 120$, the abrupt shape changes from spherical to large prolate shape are observed after the predicted shell closures at $N = 184$ and 258.
This may lead to the sudden changes in neutron rms radii in these regions, as discussed in Sec.~\ref{nrr}.

%----------------------------------------------------------------------------------------
\subsubsection{Potential energy curves}
\label{PEC}
\begin{figure*}[htbp]
  \centering  
  \includegraphics[width=\linewidth]{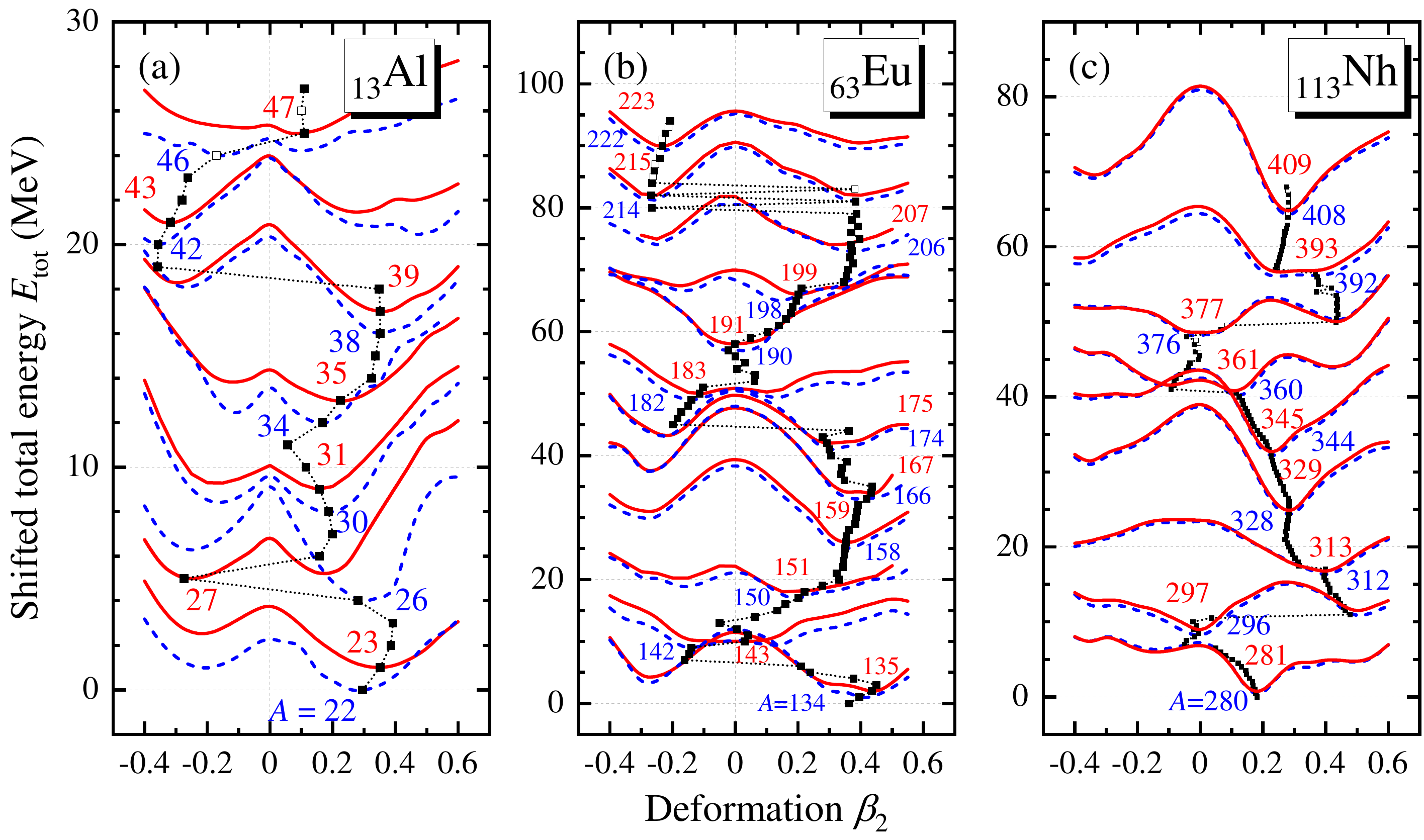}
  \caption{(Color online) Evolution of potential energy curves of $^{22,26,\cdots,46}$Al (a), $^{134,142,\cdots,222}$Eu (b), and $^{280,296,\cdots,408}$Nh (c), together with their neighboring $^{23,27,\cdots,47}$Al (a), $^{135,143,\cdots,223}$Eu (b), and $^{281,297,\cdots,409}$Nh (c) from the constrained DRHBc calculations with PC-PK1. For clarity, in each panel, the PEC of the lightest isotope ($^{22}$Al, $^{134}$Eu, and $^{280}$Nh, respectively) is renormalized to its ground state (filled square), and other PECs are shifted upward one by one by $1$ MeV for Al and Eu and by 0.5 MeV for Nh, per increasing 1 neutron. Ground-state deformations are denoted by squares.}
\label{fig18}
\end{figure*}
%----------------------------------------------------------------------------------------

To further understand the shape evolution, the PECs obtained from the constrained DRHBc calculations are presented in Fig.~\ref{fig18} for selected Al, Eu, and Nh isotopes.
The evolution of ground-state deformation in the corresponding isotopes is also depicted for guidance.

Most PECs exhibit multiple minima.
The global minima from constrained calculations are completely consistent with the ground states from unconstrained ones, which further validates the self-consistency and reliability of the DRHBc calculations.
Similar to the findings for even-$Z$ nuclei in Ref.~\cite{Guo2024mass}, the PEC of an odd-$N$ nucleus closely resembles that of its even-$N$ neighbor in most cases.

As shown in Fig.~\ref{fig18}, all Al isotopes exhibit deformed shapes in ground states. 
Even $^{33}$Al with $N=20$ retain small deformation with $\beta_2 = 0.056$, which may arise from the polarization effect of the unpaired odd proton.
The PECs of most Al isotopes have both prolate and oblate minima.
For isotopes prior to the $N = 20$ closed shell, oblate ground state can only be found in $^{27}$Al, \textit{i.e.}, the nuclear shape changes from $\beta_2 = 0.281$ in $^{26}$Al, to $\beta_2 = -0.274$ in $^{27}$Al, then back to $\beta_2 = 0.158$ in $^{28}$Al.
For isotopes with $N \geq 20$, $^{41\text{--}45}$Al are oblate and drastic deformation changes happen in $^{40}$Al and $^{45}$Al.
Such drastic deformation changes arise from the competition between the prolate and oblate minima with similar energies.
% and different orbitals of blocked proton. 
It is noted that the triaxial deformation may play an important role in Al isotopes.
The TRHBc calculations with PC-PK1 have predicted triaxial deformation in $^{26\text{--}28,30}$Al and $^{38\text{--}43,45}$Al~\cite{ZhangTRHBc2025PRC}.
The appreciable deformation in $^{38\text{--}43}$Al around $N = 28$ indicates the quenching of $N=28$ shell closure in this neutron-rich region, similarly as in Na isotopes~\cite{Zhang2023PRC}.

For the Eu isotopes, the evolutions of the deformation and PECs resemble those of neighboring even-$Z$ Gd isotopes shown in Ref.~\cite{Guo2024mass}.
The sudden shape changes from $\beta_2 = 0.362$ in $^{177}$Eu to $\beta_2 = -0.199$ in $^{178}$Eu, and from $\beta_2 = 0.387$ in $^{212}$Eu to $\beta_2 = -0.266$ in $^{213}$Eu can be found, similar as in Gd isotopes~\cite{Guo2024mass}.
These shapes changes may respectively correspond to the triaxially deformed nuclei $^{170,172,180,182,184}$Gd and the shape changes from $\beta_2 = 0.35$ in $^{210}$Gd to $\beta_2 = -0.25$ in $^{212}$Gd predicted in the triaxial relativistic Hartree-Bogoliubov calculations with PC-PK1~\cite{Yang2021PRC}. 

For the Nh isotopes, the evolutions of the deformation and PECs resemble those of neighboring even-$Z$ Fl isotopes shown in Ref.~\cite{Guo2024mass}.
There are spherical and prolate minima in the PECs of $^{296,297}$Nh and $^{376,377}$Nh.
The competition of these minima may lead to the sudden shape changes from $\beta_2 = 0.036$ in $^{301}$Nh to $\beta_2 = 0.478$ in $^{302}$Nh, and from $\beta_2 = 0.085$ in $^{379}$Nh to $\beta_2 = 0.433$ in $^{380}$Nh.

%----------------------------------------------------------------------------------------
\subsection{Neutron density distributions}\label{Ndd}
\begin{figure*}[htbp]
  \centering
  \includegraphics[width=1.0\textwidth]{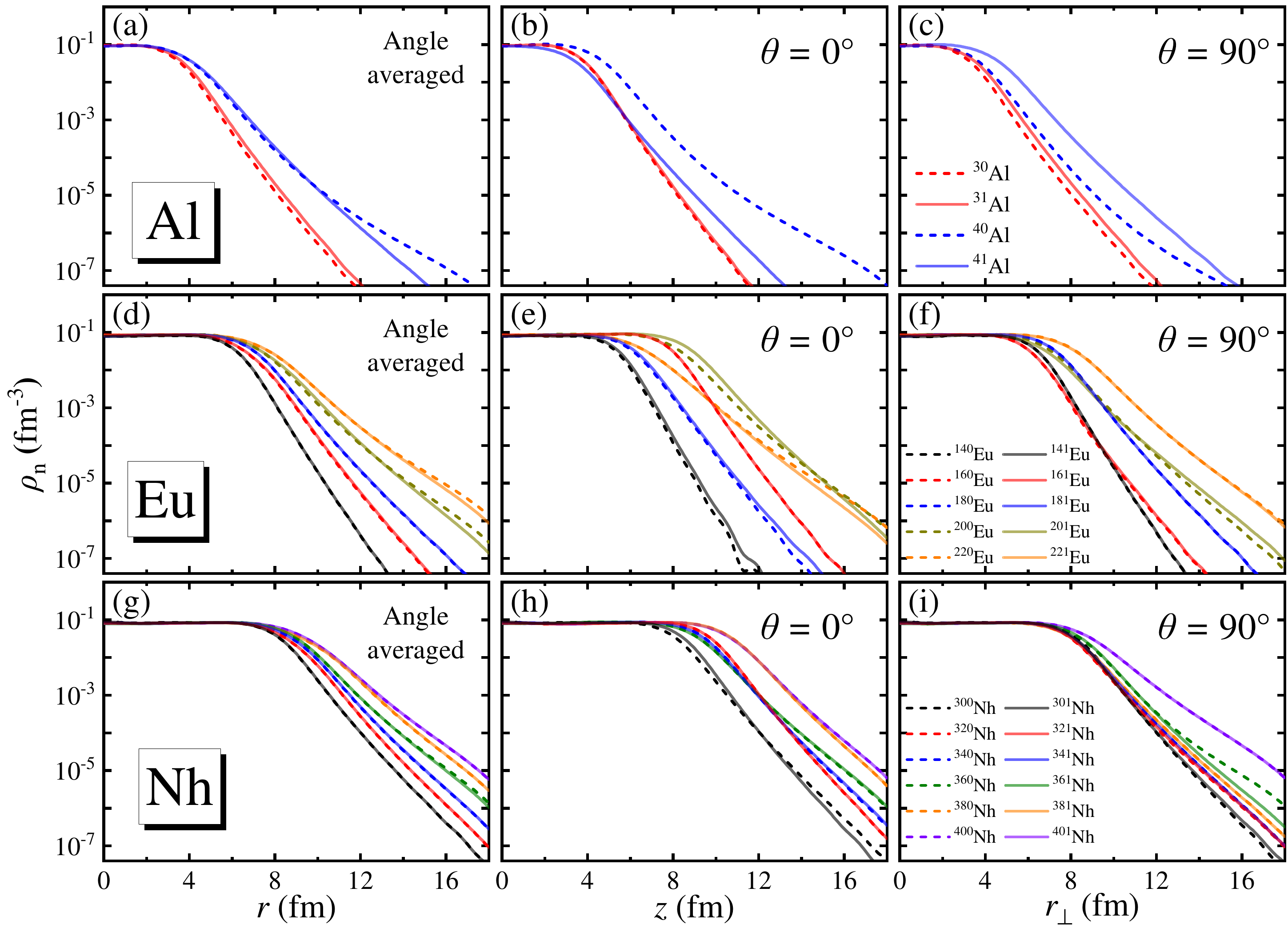}
  \caption{(Color online) Angle averaged neutron density distribution (Angle averaged), the neutron density distribution along the symmetry axis $z$ ($\theta=0^\circ$), and that perpendicular to the symmetry axis with $r_\perp = \sqrt{x^2 + y^2}$ ($\theta=90^\circ$), for selected odd-odd isotopes $^{30,40}$Al and their neighboring odd-even isotopes $^{31,41}$Al (a, b, c), $^{140,160,\cdots,220}$Eu and $^{141,161,\cdots,221}$Eu (d, e, f), and $^{300,320,\cdots,400}$Nh and $^{301,321,\cdots,401}$Nh (g, h, i) in the DRHBc calculations with PC-PK1.}
\label{fig19}
\end{figure*}
%----------------------------------------------------------------------------------------

As the continuum, deformation effects, large spatial distributions, and their coupling are treated self-consistently~\cite{Zhou2010PRC}, the DRHBc theory can adequately describe nucleon density distributions for both stable and exotic nuclei, as demonstrated for even-$Z$ nuclei in Refs.~\cite{Zhang2022mass,Guo2024mass}. 

For odd-$Z$ nuclei, Fig.~\ref{fig19} shows the neutron density profiles of selected odd-odd isotopes $^{30,40}$Al, $^{140,160,\cdots,220}$Eu, and $^{300,320,\cdots,400}$Nh, together with their corresponding odd-even isotopes $^{31,41}$Al, $^{141,161,\cdots,221}$Eu, and $^{301,321,\cdots,401}$Nh.
The angle averaged neutron density distributions, the neutron density distributions along the symmetry axis ($\theta=0^\circ$), and those perpendicular to the symmetry axis ($\theta=90^\circ$) are depicted in the left, middle, and right panels, respectively.

For the Al isotopes, the neutron density distributions in $^{40,41}$Al are more extended than those of $^{30,31}$Al, including the angle averaged distributions and those along $\theta=0^\circ$ and $90^\circ$, as shown in Figs.~\ref{fig19}(a), (b), and (c), respectively.
For the angle averaged neutron density distribution shown in Fig.~\ref{fig19}(a), the neutron density distribution of $^{31}$Al resembles that of its odd-$N$ neighbor $^{30}$Al, and this still holds for the $^{40,41}$Al for $r < 10$ fm.
Notably, the angle averaged neutron density distribution of $^{40}$Al is more extended than that of $^{41}$Al for $r > 10$ fm.
The more extended tail in $^{40}$Al originates from its blocked neutron orbital with $m^\pi = 1/2^{-}$, whose dominant component is the $2p_{1/2}$.
In Fig.~\ref{fig19}(b) for $\theta=0^\circ$, the density distribution of $^{40}$Al is more extended than that of $^{41}$Al.
In Fig.~\ref{fig19}(c) for $\theta=90^\circ$, the density distribution of $^{41}$Al is more extended than that of $^{40}$Al.
This stems from the fact that $^{40}$Al is prolate with $\beta_2 = 0.35$, and $^{41}$Al is oblate with $\beta_2 = -0.36$.

For the Eu isotopes, the angle averaged neutron density distributions in Fig.~\ref{fig19}(d) extend almost monotonically with neutron number.
The angle averaged density distributions of odd-$N$ nuclei resemble those of their even-$N$ neighbors, and this still holds for the density distributions along $\theta=0^\circ$ in Fig.~\ref{fig19}(e) and $90^\circ$ in Fig.~\ref{fig19}(f).
In Fig.~\ref{fig19}(e) for $\theta=0^\circ$, the density distributions in $^{160,161}$Eu are more extended than those in $^{180,181}$Eu, and even than those in $^{220,221}$Eu for $z < 10$ fm, which is possibly attributed to the pronounced prolate shape $\beta_2 = 0.355$ in $^{160}$Eu and $\beta_2 = 0.362$ in $^{161}$Eu.
Similarly, the density distributions for $\theta=0^\circ$ in $^{200,201}$Eu are more extended than those in $^{220,221}$Eu, which is related with the prolate shape $\beta_2 = 0.211$ in $^{200}$Eu and $\beta_2 = 0.345$ in $^{201}$Eu.
In Fig.~\ref{fig19}(f) for $\theta=90^\circ$, the density distributions in $^{140,141}$Eu are more extended than those in $^{160,161}$Eu for $r_{\perp} < 9$ fm, and the density distributions in $^{180,181}$Eu are more extended than those in $^{200,201}$Eu.
This stems from the fact that $^{140,141,180,181}$Eu are oblate while $^{160,161,200,201}$Eu are prolate.

For the Nh isotopes, the angle averaged neutron density distributions in Fig.~\ref{fig19}(g) extend almost monotonically with neutron number.
The angle averaged density distributions of odd-$N$ nuclei are similar to those of their even-$N$ neighbors, and this generally holds for the density distributions along $\theta=0^\circ$ in Fig.~\ref{fig19}(h) and $90^\circ$ in Fig.~\ref{fig19}(i), except for the difference between $^{360}$Nh and $^{361}$Nh for $ r_\perp > 13$~fm for $\theta=90^\circ$.
In Fig.~\ref{fig19}(h) for $\theta=0^\circ$, the density distributions in $^{320,321,340,341,360,361}$Nh are close to each other for $z < 12$ fm.
This may be related with the overall decreasing trend of the prolate deformation from $\beta_2 = 0.283$ in $^{320}$Nh to $\beta_2 = 0.115$ in $^{361}$Nh, as shown in Fig.~\ref{fig18}.
Similarly, the density distributions in $^{380,381}$Nh are close to those in $^{400,401}$Nh as well.
The reason may be that the prolate deformations $\beta_2 \approx 0.43$ in $^{380,381}$Nh are larger than $\beta_2 \approx 0.26$ in $^{400,401}$Nh.
In Fig.~\ref{fig19}(i) for $\theta=90^\circ$, the density distributions in $^{300,301,320,321,340,341,360,361,380,381}$Nh are quite close to each other for $r_\perp < 10$~fm, while the density distributions in $^{400,401}$Nh are significantly more extended than those of the other isotopes.

Thus, a conclusion can be drawn that the neutron density distributions manifest not only the neutron diffuseness with the increasing neutron number but also the deformation effects in odd-$Z$ nuclei, which is the same as even-$Z$ nuclei in Ref.~\cite{Guo2024mass}.

%----------------------------------------------------------------------------------------
\subsection{Neutron potential and diffuseness}
\begin{figure*}[htbp]
  \centering
  \includegraphics[width=1.0\textwidth]{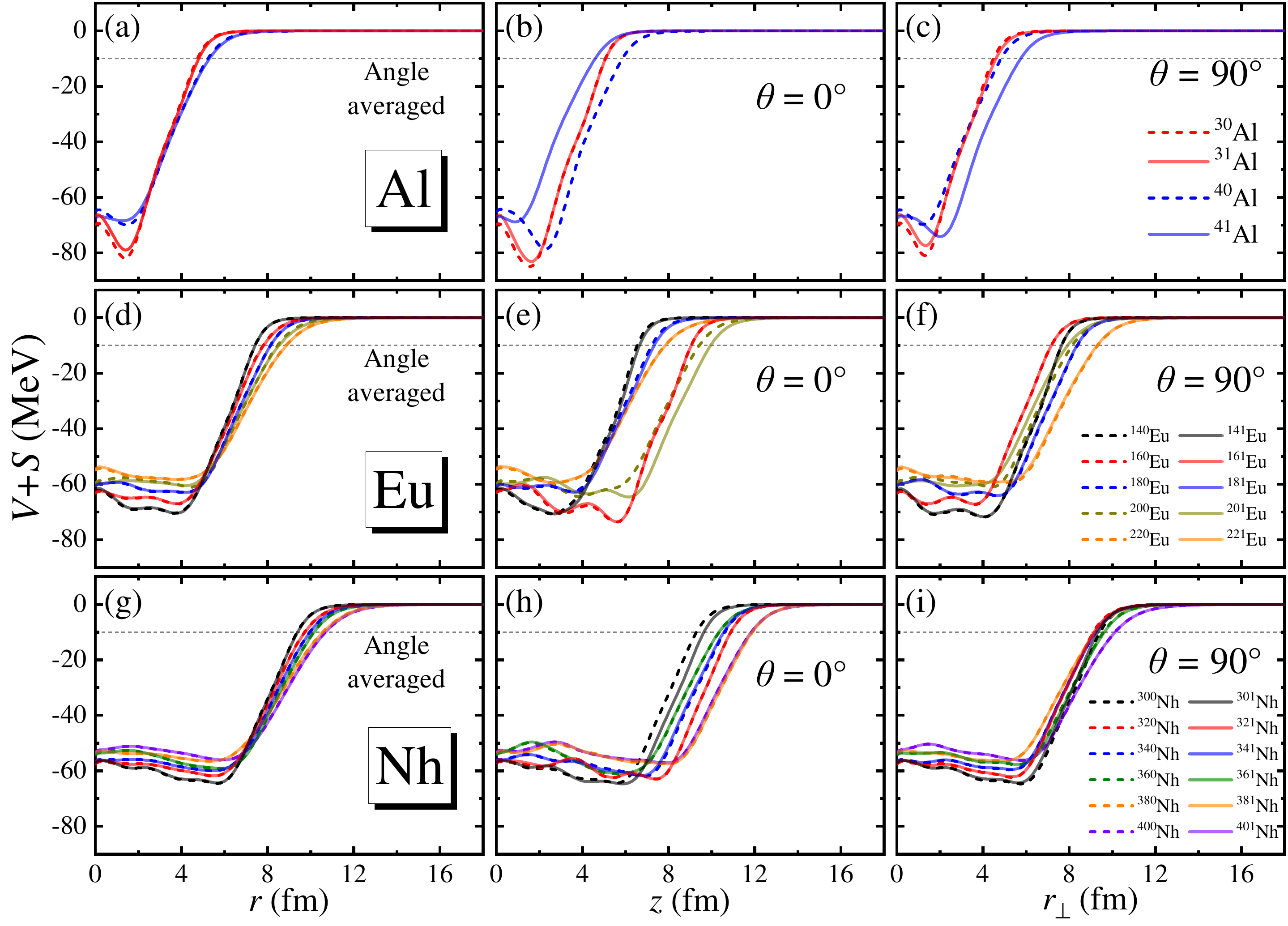}
  \caption{(Color online) Same as Fig.~\ref{fig19}, but for the neutron mean-field potential $V+S$. A gray dashed line at $-10$~MeV is plotted to demonstrate the diffuseness of the potential.}
\label{fig20}
\end{figure*}
%----------------------------------------------------------------------------------------

In DRHBc, the mean-field potential $V(\bm r) + S(\bm r)$, defined by Eqs.~\eqref{P_S} and~\eqref{P_V}, is calculated self-consistently incorporating the continuum, the deformation, and the pairing correlations. The diffuseness evolution of the potential with the neutron number~\cite{Meng1999NPA} has provided valuable guidance for the high precision mass model~\cite{Wang2014PLB}. 

The neutron mean-field potentials for selected odd-odd isotopes $^{30,40}$Al, $^{140,160,\cdots,220}$Eu, and $^{300,320,\cdots,400}$Nh, together with their corresponding odd-even isotopes $^{31,41}$Al, $^{141,161,\cdots,221}$Eu, and $^{301,321,\cdots,401}$Nh, are presented in Fig.~\ref{fig20}, in terms of the angle averaged potential and those along ($\theta=0^\circ$) and perpendicular to ($\theta=90^\circ$) the symmetry axis. To examine the diffuseness of the potential more clearly, one can compare the position where $V + S = -10$~MeV for each potential~\cite{Meng1999NPA,Meng2006PPNP}. For this purpose, a gray dashed line at $-10$~MeV is plotted.

For the Al isotopes, the angle averaged neutron mean-field potentials in Fig.~\ref{fig20}(a) for $^{30}$Al and $^{40}$Al are similar to those for $^{31}$Al and $^{41}$Al, respectively.
Compared with $^{30,31}$Al, in $^{40,41}$Al, the depth of the potentials is decreased, the surface of the potentials moves outward, and the potentials become more diffused due to the increase of neutron number.
In Fig.~\ref{fig20}(b) for $\theta=0^\circ$ and Fig.~\ref{fig20}(c) for $\theta=90^\circ$, the potential in $^{30}$Al remains similar to that in $^{31}$Al.
The potentials in $^{30,31,40}$Al are more diffused than that in $^{41}$Al for $\theta=0^\circ$, and less diffused for $\theta=90^\circ$. 
This may come from the differences in their deformations, similar to the discussion for densities in Sec.~\ref{Ndd}.

For the Eu isotopes, in Fig.~\ref{fig20}(d) for the angle averaged neutron mean-field potential, the depth decreases almost monotonically, the surface moves outward, and the diffuseness generally increases with the neutron number.
The angle averaged potentials in odd-$N$ nuclei are similar to those in their even-$N$ neighbors, and this generally holds for potentials along both $\theta=0^\circ$ in Fig.~\ref{fig20}(e) and $\theta=90^\circ$ in Fig.~\ref{fig20}(f), except for the slight difference of the $\theta=0^\circ$ potential in $^{200}$Eu and $^{201}$Nh with $5 < z < 12$~fm.
In Fig.~\ref{fig20}(e) for $\theta=0^\circ$, the potentials in $^{160,161,200,201}$Eu are much more diffused than the others, while those for $\theta=90^\circ$ in Fig.~\ref{fig20}(f) are less diffused than those in $^{180,181,220,221}$Eu, as guided by the $V + S = -10$~MeV gray dashed line.
This is because $^{160,161,200,201}$Eu are prolate while other isotopes are oblate.

For the Nh isotopes, in Fig.~\ref{fig20}(g) for the angle averaged neutron mean-field potential, the depth decreases almost monotonically, the surface moves outward, and the diffuseness generally increases with the neutron number.
The angle averaged potentials in odd-$N$ nuclei are similar to those in their even-$N$ neighbors, and this generally holds for potentials along both $\theta=0^\circ$ in Fig.~\ref{fig20}(h) and $\theta=90^\circ$ in Fig.~\ref{fig20}(i), except for the slight difference of the $\theta=0^\circ$ potential in $^{300}$Nh and $^{301}$Nh with $6 < z < 10$~fm.
In Fig.~\ref{fig20}(h) for $\theta=0^\circ$, the potentials in $^{320,321,340,341,360,361}$Nh become less diffused with increasing neutron number and the potentials in $^{380,381}$Nh are very close to those in $^{400,401}$Nh, as guided by the $V + S = -10$~MeV gray dashed line.
% which are clearly manifested in the shift of the positions where $V + S = -10$~MeV. 
In Fig.~\ref{fig20}(i) for $\theta=90^\circ$, the potentials in $^{300,301,320,321,340,341,360,361,380,381}$Nh are quite close to each other.
These may come from the overall decreasing trend of their prolate deformations, similar to the discussion for densities in Sec.~\ref{Ndd}.

%%%%%%%%%%%%%%%%%%%%%%%%%%%%%%%%%%%%%%%%%%%%%%%%%%%%%%%%%%
%                    begin  summary
%%%%%%%%%%%%%%%%%%%%%%%%%%%%%%%%%%%%%%%%%%%%%%%%%%%%%%%%%%

%----------------------------------------------------------------------------------------
\section{Summary}\label{summary}
%----------------------------------------------------------------------------------------

In summary, we have constructed the nuclear mass table with the effects of pairing, deformation, and continuum incorporated in a self-consistent and microscopic way for all nuclei with $8 \le Z \le 120$ by using the DRHBc theory with the density functional PC-PK1, extending and completing the previous works on even-even~\cite{Zhang2022mass} and even-$Z$~\cite{Guo2024mass} nuclei.
The calculated binding energies, two-nucleon and one-nucleon separation energies, rms radii of neutron, proton, matter, and charge distributions, quadrupole deformations, neutron and proton Fermi surfaces, and neutron and proton blocked orbitals when applicable, are tabulated.
% Systematic studies have been performed for all nuclei from the proton drip line to the neutron drip line in the DRHBc mass table. 

In total, 9495 nuclei with $8 \le Z \le 120$ are predicted to be bound in the DRHBc mass table.
The rms deviation for binding energies from the 2380 available experimental data is $1.444$ MeV, placing this work among the most accurate density functional descriptions of nuclear masses to date.
The extracted nucleon separation energies and their evolution across the nuclear landscape are presented and discussed.
The drastic changes at magic numbers in nucleon separation energies are discussed as well, revealing traditional shell closures and potential new ones.
The weakly bound nuclei with nucleon separation energies less than 1~MeV are discussed.
% The traditional magic numbers are well reproduced and new magic numbers $N=184$ and $258$ are predicted.
% The odd-even staggering exhibited in the one-neutron and one-proton separation energies is examined.

The accuracy of the nuclear masses and nucleon separation energies and the predicted limits of the nuclear landscape are compared with the results from other mass models.
The predicted proton drip line is generally consistent with experimental observations, and close to those predicted by other models.
The predicted neutron drip line is generally the most extended in comparison with other models, mainly because of the proper treatment of the continuum and the adopted density functional.
Several peninsulas of stability are obtained near the neutron drip line at $Z \approx 60, 88$, and 110, which are mainly attributed to the effects of deformation.
 % and influenced by the pairing correlations and continuum in a self-consistent way.

The pairing gaps, the pairing energies, and the two-nucleon gaps are extracted for both neutron and proton.
Based on the vanishing pairing energies and pronounced two-nucleon gaps, the traditional magic numbers are well reproduced, and new neutron magic numbers $N=184$ and $258$ are predicted.
Similarly, the pronounced two-neutron gaps at $N=40$ and $162$ for several isotones indicate possible shell closures. 
The pronounced two-proton gaps also appear in the nuclei around $Z = 14$ and along the $N \sim Z$ band.
% and a proton subshell with $Z=14$ for several Si isotopes. 
The pronounced pairing energies and vanishing two-neutron gaps indicate the quenching of traditional neutron shell closures at $N = 28, 50$, and $82$ near the neutron drip lines. 

The $\alpha$ decay energies are extracted and compared.
With the inclusion of deformation effects, the rms deviation from 1090 available experimental data is $\sigma(Q_\alpha) = 0.841$~MeV, which is significantly improved from $\sigma(Q_\alpha) \approx 2$ MeV in the RCHB results~\cite{Zhang2016CPC}. 
Future work is expected to improve the description of $Q_\alpha$ and to estimate systematically the $\alpha$ decay half-lives.
% It is interesting to estimate the $\alpha$ decay half-lives by combining the DRHBc theory with the WKB method and further improve the description by the collective Hamiltonian method.

For the 1014 nuclei with charge radius measured, the rms deviation between the calculated results and the data is $\sigma(R_\mathrm{ch}) = 0.035$ fm.
The systematic trend of the neutron rms radii for nuclei with $8 \le Z \le 120$ generally follows the empirical formula $R_\mathrm{n} \propto N^{1/3}$ with a few exceptions.
The pronounced deviations for some extremely neutron-rich nuclei are possible signals for the halo or giant halo phenomena.
The drastic increases after the predicted shell closures at $N = 184$ and 258 can be attributed to the deformation effects.

The quadrupole deformations for nuclei with $8 \leq Z \leq 120$ and the PECs for selected isotopes are presented and discussed.
The weakening or even collapse of the traditional shell closures is further demonstrated by the deformation deviating from the spherical shape.
The drastic shape changes, \textit{e.g.}, those around $N = 115$ with $60\leq Z \leq 71$, come from the competition between different minima with comparable energies.
The possible triaxial deformations and shape coexistence in those regions are discussed with reference to the results in calculations including triaxial deformations.
The shape evolution in Al, Eu, and Nh isotopic chains is further understood from PECs in constrained DRHBc calculations.

Finally, the angle averaged neutron density distributions and neutron mean-field potentials as well as those along and perpendicular to the symmetry axis for selected odd-odd Al, Eu, and Nh isotopes and their corresponding neighbors are presented.
The diffuseness of the neutron density distributions and mean-field potentials generally increases with the increase of neutron number.
The deformation effects are also manifested in the evolutions of the density distributions and mean-field potentials.

The successful construction of the DRHBc mass table with relativistic density functional PC-PK1 provides a comprehensive and microscopic description for nuclei with $8\le Z\le120$. 
Based on the DRHBc results, future work will be devoted to predict nuclear level densities, neutron capture rates, and $\beta$-decay lifetimes to establish a microscopic and self-consistent database for nuclear astrophysics.

% The comprehensive DRHBc mass table has been constructed, with the ground-state properties of nuclei with $8\le Z\le120$ included.
% In the future, the descriptions of ground-state properties are expected to be further improved by incorporating beyond-mean-field correlations with collective Hamiltonian method.
% A systematic search for possible halo candidates could be carried out, providing useful guidance for experimentalists.
% Further extensions of the DRHBc theory are also in progress, \textit{e.g.}, including the triaxial deformation degree of freedom~\cite{Zhang2023PRCL} and accounting for the current and time-odd field~\cite{Pan2024PLB}.

\begin{acknowledgments}
This work was partly supported by the National Natural Science Foundation of China (Grants No.~12435006, No.~12475117, No.~11775099, No.~12075085, No.~12047568, No.~12205308, No.~12265012, No.~12275082, No.~12375119, No.~11975041, No.~12047503, No.~11975237, No.~12305125, No.~11961141004, No.~12375109, No.~12447101, No.~12375118, No.~12435008, No.~W2412043, No.~12575130, No.~12375117, No.~12605194, and No.~12675160), the National Key R\&D Program of China (Contracts No.~2024YFE0109803, No.~2024YFA1612600, No.~2020YFA0406001, No.~2020YFA0406002, No.~2021YFA1601500, No.~2022YFA1602401, No.~2023YFA1606500 and No.~2024YFE0109800), the State Key Laboratory of Nuclear Physics and Technology, Peking University (Grant No.~NPT2023KFY02), the Strategic Priority Research Program of Chinese Academy of Sciences (Grants No.~XDB0920000 and No.~XDB1550000), the National Key Laboratory of Neutron Science and Technology (Grants No.~NST202401016 and No.~NST20250101), the National Key Laboratory of Nuclear Data (Grant No.~JCKY2026201C153), Guangdong Basic and Applied Basic Research Foundation (2023A1515010936), the High-performance Computing Platform of Peking University, the High-performance Computing Platform of Anhui University, the High Performance Computing Resources in the Institute for Basic Science Research Solution Center, the National Research Foundation of Korea (NRF) funded by Ministry of Science and ICT (2013M7A1A1075764), and the Fundamental Research Funds for the Central Universities (JUSRP202406002).
This work was partly supported by the National Research Foundation of Korea (Grant Nos.~NRF-2018R1D1A1B05048026, NRF-2020R1A2C3006177, NRF-2020K1A3A7A09080134, NRF-2021R1F1A1060066, NRF-2021R1A6A1A03043957, NRF-2023R1A2C1005398, and TOPTIER RS-2024-00436392) and the National Supercomputing Center with supercomputing resources including technical support (KSC-2020-CRE-0329, KSC-2021-CRE-0126, KSC-2021-CRE-0272, KSC-2022-CRE-0333, KSC-2023-CRE-0170, and KSC-2023-CRE-0521).
Y.K. was supported in part by the Institute for Basic Science (IBS-R031-D1).
Y.C. and C.H.L. were supported by the National Research Foundation of Korea (NRF) grant funded by the Korean government (No.~RS-2023-NR076639).
M.-H.M. was supported by the National Research Foundation of Korea(NRF), funded by the Korean government (MSIT), under Grant Nos.~RS-2026-25487837 and RS-2018-NR031074.
This work was partly supported by the Japan Society for the Promotion of Science (JSPS) Grant-in-Aid under Grants Nos.~26K07063 and 26K01431.
The results from ITP side are obtained on the High-performance Computing Cluster of ITP-CAS and the ScGrid of the Supercomputing Center, Computer Network Information Center of Chinese Academy of Sciences.
We acknowledge the hospitality at APCTP where part of this work was done.
\end{acknowledgments}

\newpage
\section*{Explanation of Tables}
\textbf{Table II. Ground-state properties of nuclei calculated by the DRHBc theory}

% \begin{tabular}{@{}p{0.0in}p{1.5in}p{4.5in}@{}}
\begin{tabular}{p{0.0in}p{1.5in}p{4.5in}}
%& & When given, the italicized numbers refer to the
%     uncertainties in the last digits of the quoted values.\\
& $Z$ & Proton number\\
& $N$ & Neutron number\\
& $A$ & Mass number\\
& $E^{\mathrm{cal}}_{\mathrm{b}}$ & Binding energy from DRHBc calculations\\
& $E^{\mathrm{cal}}_{\mathrm{b}+\mathrm{rot}}$ & Binding energy plus rotational correction energy from DRHBc calculations, which is suggested to be compared with the experimental value\\
& $E^{\mathrm{exp}}_{\mathrm{b}}$ & Binding energy from experimental data~\cite{AME2020(3)}\\
& $S_\mathrm{2n}$ & Two-neutron separation energy\\
& $S_\mathrm{2p}$ & Two-proton separation energy\\
& $S_\mathrm{n}$ & One-neutron separation energy\\
& $S_\mathrm{p}$ & One-proton separation energy\\
& $R_\mathrm{n}$ & Neutron root-mean-square radius \\
& $R_\mathrm{p}$ & Proton root-mean-square radius \\
& $R_{m}$ & Matter root-mean-square radius \\
& $R^{\mathrm{cal}}_{\mathrm{ch}}$ & Charge radius from DRHBc calculations \\
& $R^{\mathrm{exp}}_{\mathrm{ch}}$ & Charge radius from experimental data~\cite{Angeli2013ADNDT,Li2021ADNDT}\\
& $\beta_\mathrm{2n}$ & Neutron quadrupole deformation \\
& $\beta_\mathrm{2p}$ & Proton quadrupole deformation \\
& $\beta_{2}$ & Matter quadrupole deformation \\
& $\lambda_\mathrm{n}$ & Neutron Fermi surface\\
& $\lambda_\mathrm{p}$ & Proton Fermi surface\\
& $m^\pi(N)$ & Quantum numbers $m^\pi$ of the blocked neutron orbital \\
& $m^\pi(P)$ & Quantum numbers $m^\pi$ of the blocked proton orbital \\
& $\sigma$ & Rms deviations for binding energies and charge radii for each isotopic chain \\
\\

\end{tabular}

\textbf{Note: \\
* denotes the nucleus with high density instability. Since PC-PK1 is a non-linear density functional, it encounters high density instability in $^{14}$O, $^{24,25,26}$Mg, $^{25}$Al, and $^{26,28}$Si, similar to that in $^{12}$C by NL1~\cite{Reinhard1988ZPA}.\\
$\dag$ denotes the nucleus which has positive $S_{2n}$, $S_{2p}$, $S_{n}$, and $S_{p}$ as well as negative $\lambda_n$ and $\lambda_p$, but is unbound against multi-nucleon emission.\\
When the pairing energy vanishes, the Fermi surface is chosen to be the energy of the last occupied single-particle state.}

\newpage
% \begin{turnpage}
\begin{landscape}
\pagestyle{empty}
\centering

\scriptsize 
\setlength{\tabcolsep}{1.8pt}

%
\newpage
\end{landscape}
% \end{turnpage}
%%------------------------------------------------------------------------------------------
% \bibliography{myrefs}
\end{CJK*}
\end{document}